\documentclass[review]{elsarticle}
\usepackage{lineno,hyperref}
\usepackage{verbatim} 
\usepackage{epsfig}
\usepackage{amssymb} 
\usepackage{mathtools}
\usepackage{lineno}
\usepackage{tikz}        
\usepackage{etoolbox}
\usepackage{ulem}         
\usepackage[caption=false]{subfig}
\usepackage{textcomp}
\usepackage{float}
\usepackage{caption} 
\usepackage{booktabs}
\usepackage{url}
\restylefloat{table}

\begin{document}

\begin{frontmatter}

\title{Faster and safer evacuations induced by closed 
vestibules}

\author[add1]{I.M.~Sticco}
 \address[add1]{Universidad de Buenos Aires, Facultad 
de Ciencias Exactas y Naturales, Departamento de F\'\i sica, Buenos Aires, 
Argentina.}

 \author[add2]{G.A.~Frank}
 \address[add2]{Unidad de Investigaci\'on y Desarrollo de las 
Ingenier\'\i as, Universidad Tecnol\'ogica Nacional, Facultad Regional Buenos 
Aires, Av. Medrano 951, 1179 Buenos Aires, Argentina.}

\author[add1,add3]{C.O.~Dorso\corref{cor1}}%

 \address[add3]{CONICET-Universidad de Buenos Aires, IFIBA, Buenos Aires, 
Argentina.}

\begin{abstract}
Improving emergency evacuations is a top priority in human safety and in 
pedestrian dynamics. In this paper, we use the 
social force model, in order to optimize high-anxiety pedestrian evacuations. We 
explore two architectural layouts, the 1-door vestibule, and the 2-doors 
vestibule. The ``vestibule'' is defined as the room next to the exit door and it 
is characterized by two structural parameters: the vestibule width ($d$) and 
the vestibule door width ($w$). We found that, specific values of $d$ and 
$w$, can almost double the evacuation flow compared to the no-vestibule 
scenario. The key to this achievement is that the density (close to 
the exit door) can be controlled by $d$ and $w$. Therefore, it is possible to 
tune these parameters to a density that maximizes the available space 
while preventing the formation of blocking clusters at the exit door ($\rho \sim 
2.5\,$p/m$^2$). 
As opposed to the optimal condition, low-density values ($\rho \sim 
1\,$p/m$^2$) lead to suboptimal flow since there is unused space left; while 
higher density values ($\rho \sim 4\,$p/m$^2$) also lead to suboptimal flow due 
to the presence of blocking clusters at the exit. Moreover, we take into 
account the usually foreseen fact that high pressures can actually be reached 
at the exit, threatening the health of pedestrians. Therefore, we 
studied the crowd pressure using the agents' overlap as an indicative. We found 
that the explored vestibules reduce the crowd pressure compared to the 
no-vestibule situation. In particular, we show that the 2-doors vestibule 
scenario performs better than the 1-door vestibule, because it reduces the 
overall local density (by enforcing the crowd to spread out more).
\end{abstract}


\end{frontmatter}


\section{Introduction}

Improving emergency evacuations is an issue of fundamental importance for human 
safety. Despite the efforts being done by crowd management legislations 
worldwide, catastrophic events such as human stampedes and fatal emergency 
evacuations are still common~\cite{daniel2022crowd,de2019human}. In the last 
years, the scientific community has arrived at a growing number of ideas to 
address this problem~\cite{helbing2000simulating,kirchner2003friction}. The 
proposed solutions can be classified into three 
categories~\cite{haghani2020optimising}: planning-based solutions, 
behavioral-based solutions and architectural-based solutions.\\

In the first place, planning-based solutions consist of optimizing the 
departure schedule and 
planning the right path to escape a 
building~\cite{abdelghany2014modeling,pursals2009optimal}. Sometimes these 
solutions require a central authority to control the crowd. Secondly, 
behavioral-based solutions consist of proposing modifications in the attitude of 
the pedestrians~\cite{song2016selfishness,cheng2018emergence}. 
These solutions require training and proper instructions to be acquired by 
potential evacuating individuals.\\

Finally, architectural-based solutions consist of proposing design and 
infrastructure 
adjustments that reduce the evacuation time and increase safety during the 
evacuation process. Unlike the aforementioned approaches, the 
architectural-based solutions do not focus on the guidance of pedestrians nor 
their specific training.\\

Among the three approaches, the most popular one is the architectural-based. In 
particular, positioning an exit in the corner of a room~\cite{shi2019examining} 
and placing an obstacle in front of the exit door~\cite{helbing2005self} appear 
to reduce the evacuation time in both numerical 
simulations~\cite{frank2011room,sticco2022improving}, and controlled 
experiments~\cite{zhao2020experimental}. \\

In the seminal work of Helbing et al., it was proposed for the first time the 
use of a column-like obstacle to improve the evacuation 
performance~\cite{helbing2000simulating}. Since then, 
many efforts have been made to understand this phenomenon. Although some results 
favor the initial hypothesis~\cite{helbing2005self}, some others challenge the 
statement that the column-like obstacle can enhance the evacuation 
performance~\cite{garcimartin2018redefining}.\\

Placing panel-like obstacles instead of pillar-like obstacles, although it is a 
simpler proposal, it appears as a more promising solution in terms of increasing 
the evacuation flow. There are 
numerical results~\cite{frank2011room} as well as experimental 
evidence~\cite{zhao2020experimental} that support this idea.\\

The authors in Ref.~\cite{sticco2022improving} postulate the idea that placing two 
panel-like obstacles in front of the exit door improves the evacuation performance. 
They introduce the concept of ``vestibule'' (say, the space near the exit door) 
as a more relevant one than the obstacle itself. The authors found that the 
evacuation flow is determined by the density at the vestibule. Moreover, the 
density can be controlled by architectural features such as the number of 
vestibule doors, the vestibule width and the wall friction coefficient.\\

In addition to improving the evacuation flow, it is important to reduce the 
physical pressure exerted on individuals. Many different 
approaches appeared in the literature regarding the pressure in 
crowd dynamics~\cite{feliciani2020systematic}. The pressure used in some 
empirical and 
experimental studies is defined as the product of the local density and the 
velocity variance~\cite{helbing2007dynamics}. This 
magnitude has proved to correlate with the probability 
of tripping and the anxiety level~\cite{garcimartin2017pedestrian}.\\

Other approach worth mentioning is the numerical study done by Cornes et 
al. which define pressure as a function of the normal 
forces acting on the 
individuals~\cite{cornes2017high}. In line with this, it is worth remarking the 
experimental 
efforts done by different authors who craft ``pressure vests'' in order to 
measure the contact forces applied to the human thorax~\cite{wang2018study}.\\

This paper is a numerical study that is framed in the architectural-based 
approach. We investigate emergency evacuations in the presence of a closed 
vestibule. The closed vestibule is defined as the room next to the exit door 
which is enclosed by panel-like obstacles. We show that this 
layout is capable of improving emergency evacuations for two reasons. On 
one hand, it substantially increases the evacuation flow of pedestrians. On the 
other hand, it reduces the pressure exerted on them.\\

The paper is organized as follows. In Section \ref{background}, we present the 
model and theoretical definitions. In section \ref{numerical_simulations}, we 
describe the numerical simulations and the explored layouts. The results are 
exhibited in Section \ref{results}. We finally resume the 
conclusions in Section \ref{conclusions}.\\

\section{\label{background} Background}

\subsection{\label{sfm}The Social Force Model}

The social force model~\cite{helbing2000simulating} provides the necessary 
framework for simulating  the collective dynamics of pedestrians (\textit{i.e.} 
self-driven agents). The pedestrians are represented as moving particles that 
evolve according to the presence of either ``socio-psychological'' forces 
and physical forces. The equation of motion for any agent $i$ of mass $m_i$ 
reads

\begin{equation}
 m_i\,\displaystyle\frac{d\mathbf{v}_i}{dt}=\mathbf{f}_d^{(i)}+
 \displaystyle\sum_{j=1}^N\mathbf{f}_s^{(ij)}+
 \displaystyle\sum_{j=1}^N\mathbf{f}_p^{(ij)}\label{newton_ec}
\end{equation}

\noindent where the subscript $j$ corresponds to any neighboring agent or
the walls. The three forces $\mathbf{f}_d$, $\mathbf{f}_s$ and $\mathbf{f}_p$
are different in nature. The desire force $\mathbf{f}_d$ represents the
acceleration of a pedestrian due to his/her own will.  The
social force $\mathbf{f}_s$, instead, describes the tendency of the  pedestrians
to stay away from each other. The physical force 
$\mathbf{f}_p$  stands for both the sliding friction and the repulsive body 
force. \\

The pedestrians' own will is modeled by the desire force $\mathbf{f}_d$.  This
force stands for the acceleration required to move  at the
desired walking speed $v_d$. For a fixed parameter $\tau$ reflecting the 
reaction time, the desire force is modeled as follows

\begin{equation}
\mathbf{f}_d^{(i)}=m\,\displaystyle\frac{v_d^{(i)}\,
\hat{\mathbf{e}}_d^{(i)}(t)-
 \mathbf{v}^{(i)}(t)}{\tau}
\end{equation}

\noindent where $\hat{\mathbf{e}}(t)$ represents the unit vector pointing to the
target position and $\mathbf{v}(t)$ stands for the agent velocity at time $t$.\\

The tendency of any individual to preserve his/her personal space is
accomplished by the social force $\mathbf{f}_s$. This force is expected to
prevent the agents from getting too close to each other (or to the 
walls) in any usual environment. The model for this kind of  
``socio-psychological''
behavior is as follows

\begin{equation}
 \mathbf{f}_s^{(i)}=A\,e^{(R_{ij}-r_{ij})/B}\,\hat{\mathbf{n}}_{ij}
 \label{eqn_social}
\end{equation}

\noindent where $r_{ij}$ is the distance between the centers of mass of 
particles $i$ and $j$, 
and $R_{ij}=R_i+R_j$ is the sum of the pedestrians
radius. The unit vector $\hat{\mathbf{n}}_{ij}$ points from pedestrian $j$ to
pedestrian $i$, meaning a repulsive interaction. The parameter $B$ is a 
characteristic scale that plays the role of a fall-off length within the social 
repulsion. At the same time, the parameter $A$ represents the intensity of the 
social repulsion. \\ 

The  expression for the physical force (the friction force plus the body 
force) has been inspired from the granular  matter 
field~\cite{risto1994density}. The mathematical expression reads as follows

\begin{equation}
 \mathbf{f}_p^{(ij)}=\kappa_t\,g(R_{ij}-r_{ij})\,
(\Delta\mathbf{v}^{(ij)}\cdot\hat{\mathbf{t}}_{ij})\,\hat{\mathbf{t}}_{ij}+
k_n\,g(R_{ij}-r_{ij})\,
\,\hat{\mathbf{n}}_{ij}\label{eqn_friction}
\end{equation}

\noindent where $g(R_{ij}-r_{ij})$ equals $R_{ij}-r_{ij}$ if $R_{ij}>r_{ij}$ and
vanishes otherwise. $\Delta\mathbf{v}^{(ij)}\cdot\hat{\mathbf{t}}_{ij}$
represents the relative tangential velocities of the sliding  bodies (or between
the pedestrian and the walls).    \\

The sliding friction occurs in the tangential direction while the body force
occurs in the normal direction. Both are assumed to be linear with respect to
the net distance between contacting agents. The sliding friction is also
linearly related to the difference between the tangential velocities.\\

The coefficients $\kappa_t$ (for the sliding friction) and $k_n$ (for the  body 
force) are related to the contacting surface materials and the 
body stiffness, respectively. The friction force between an agent and a 
wall/panel has the same mathematical expression as the friction between two 
agents.\\ 

The model parameter values were chosen to be the same as the 
best-fitting parameters reported in the recent study from 
\cite{sticco2021social}. These values were obtained by fitting 
the model to a real-life event that resembles an emergency 
evacuation. The parameter values are: $A=2000\,$N, $B=0.08\,$m, 
$\kappa_t=3.05\times10^{5}\,$kg/(m.s), $k_n=3600\,$N/m, $\tau=0.5\,$s.\\

\subsection{\label{bc} Blocking clusters}

A characteristic feature of pedestrian dynamics is the formation of clusters.
Clusters of pedestrians can  be defined as the set of individuals that for any
member of the group (say, $i$) there exists at least another member belonging to
the same group ($j$) in contact with the former.  Thus, we define a ``granular
cluster'' ($C_g$) following the mathematical formula given 
in \cite{strachan1997fragment}

\begin{equation}
C_g:P_i~\epsilon~ C_g \Leftrightarrow \exists~ j~\epsilon~C_g / r_{ij} 
< (R_i+R_j) \label{ec-cluster}
\end{equation}

where ($P_i$) indicate the \textit{ith} pedestrian and $R_i$ is his/her radius
(half of the shoulder-to-shoulder width). That means, $C_g$ is a set of 
pedestrians that interact not only with the social force, but also with physical 
forces (\textit{i.e.} friction force and body force). A ``blocking cluster'' is 
defined as the minimal granular cluster which is closest to the door whose 
first and last agents are in contact with the walls at both sides of the 
door (\textit{i.e.} the first and last agents are the closest 
ones to each doorjamb)~\cite{parisi2005microscopic}. Previous 
studies demonstrated that the blocking clusters play a crucial role in 
preventing 
pedestrians from getting through a 
door~\cite{parisi2005microscopic,sticco2017room,cornesmicroscopic}.

\section{\label{numerical_simulations}Numerical simulations}

We performed numerical simulations of 
pedestrians evacuating a room in the presence of a 1-door vestibule and 
a 2-doors vestibule (see Fig.~\ref{cv_1door} and Fig.~\ref{cv_2doors}, 
respectively). We simulated 
crowds of N=200 agents whose trajectories followed the equation 
of motion (\ref{newton_ec}). We used the circular specification of the social 
force model for the interaction 
forces acting on the agents (see Section \ref{sfm} for the corresponding 
mathematical expressions). The model parameter values were chosen to be the same 
as the best-fitting parameters reported in the 
recent study~\cite{sticco2021social}, and the ones cited in Section 
\ref{sfm}.\\ 

The mass of the agents was fixed at $m=80\,$kg, and the radius was set to 
$R=0.23\,$m according to data reported in \cite{littlefield2008metric}. The 
desired 
velocity (which is the parameter that controls the anxiety level) was varied in 
the interval $1\,$m/s\,$\leq v_d\leq6\,$m/s. Although the upper limit 
($v_d=6\,$m/s) may seem rather extreme, recent empirical measurements at San 
Ferm\'{i}n bull-running festival in Pamplona shows that non-professional runners 
can achieve velocities $v\sim 6\,$m/s~\cite{parisi2021pedestrian}.\\

Initially, the agents are placed in a $20$\,m\,$\times \, 
20\,$m area with random positions and velocities. The direction of the desired 
velocity was such that the agents located outside the vestibule point to the 
nearest vestibule door. Once they are inside the vestibule, the updated
target becomes the exit door. \\

The exit door's size was $DS=1.84\,$m, which is equivalent to four agents' 
diameter in accordance to \cite{sticco2021social}. The 1-door vestibule 
(Fig.~\ref{cv_1door}) is composed of 2 panel-like obstacles (\textit{i.e.} 
walls). The walls that define the vestibule extend to the side of the 
room. The door of the vestibule is placed symmetrically with respect to the 
exit door. The width of the vestibule door is $w$. See Section \ref{layouts} 
for more details.\\

The 2-doors vestibule layout (Fig.~\ref{cv_2doors}) is composed of 3 panel-like 
obstacles. At this instance, we chose a commonly expected configuration: the 
middle panel shares the same length as the exit door $DS$, and it is 
placed right in front of it. The left and right panels are as long as the room. 
Both vestibule doors have the same width $w/2$ (therefore, the total 
vestibule entry width is $w$). The middle panel is fixed (in length and 
location) regardless of $w$. This means that increasing $w$ widens each door 
outwards.\\

In this research, we analyzed the effect of varying 3 parameters: the desired 
velocity $v_d$, the vestibule width $d$, and the total vestibule door width $w$. 
For each parameter configuration, we performed 30 evacuation processes that 
finished when 90\% of agents left the room. No re-entry of agents was 
allowed.\\

We recorded the positions and velocities every 0.5\,s from the 
beginning of the simulation until the end (\textit{i.e.} when  90\% of the 
individuals have left the room). Then, we computed the mean density, 
the evacuation flow, and the probability of blocking clusters formation.\\

The density was measured in the inner vestibule region and it was defined 
as the number of pedestrians divided by the inner vestibule area (see 
Fig.\ref{cv_1door} and Fig.\ref{cv_2doors}). The area of the inner vestibule is 
$DS \times d$ (where $DS$ stands for the size of the exit door and $d$ 
is the vestibule width). The density was calculated every 0.5\,s. 
Afterward, the mean and standard deviation were computed from 
the recorded values.\\

The evacuation flow is defined as follows,

\begin{equation}
J=\frac{n}{t_e}
\label{J}
\end{equation}

\noindent where $n$ is the number of evacuated pedestrians and $t_e$ is the time it 
takes for those pedestrians to evacuate the room. The evacuation finished when 
90\% of the initial number of pedestrians abandoned the room. \\

The blocking clusters are the set of pedestrians that block a 
door (see Section \ref{bc} for a formal definition). We measured the probability 
of blocking clusters formation in three different regions: the exit door, the 
doors of the vestibule, and also the blocking clusters inside the vestibule. The 
probability is defined as the fraction of time that the blocking clusters are 
formed over the total evacuation time. \\

The numerical simulations were carried out using LAMMPS, which is a molecular 
dynamics open-access software~\cite{plimpton1995fast}. The implementation also 
required customized modules developed in C\raisebox{1mm}{\tiny ++}. The 
integration of the agents' trajectory was computed using the Velocity Verlet 
algorithm with a time-step $\Delta t=10^{-4}\,$s.\\

\subsection{\label{layouts}The explored layouts}

\begin{figure}[!htbp]
\centering
\textbf{1-door vestibule}\par\medskip
{\includegraphics[width=0.7\columnwidth]
{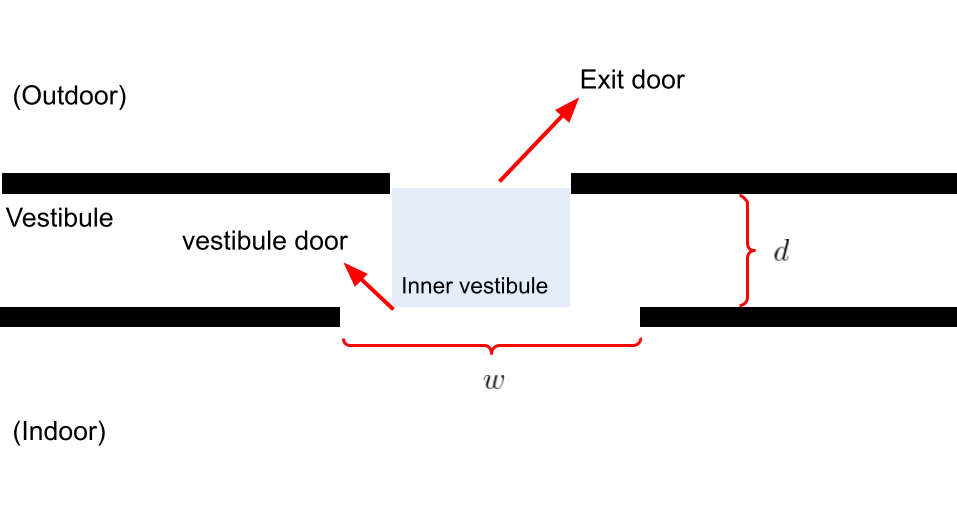}}\
\caption[width=0.49\columnwidth]{1-door vestibule layout. The vestibule door of 
width $w$ is placed in front of the exit door. The vestibule looks like 
a corridor as long as the room and width $d$. The sides of the room are not 
drawn. }
\label{cv_1door}
\end{figure}

In architecture, a vestibule is defined as a small room that leads to a larger 
space~\cite{harris2006dictionary}. In this paper, we consider vestibules placed 
next to the exit door. Two different layouts were explored: the 1-door 
vestibule and the 2-doors vestibule. Fig.~\ref{cv_1door} illustrates the 1-door 
vestibule layout. It consists of 4 walls in total. Two walls delimit the exit 
door and, the other two delimit the vestibule door.\\

The vestibule is a corridor as long as the room and width $d$. 
It has a single door of width $w$ in front of the exit door. The shady area in 
Fig.~\ref{cv_1door} represents the ``inner vestibule'', which we define 
as the rectangular area enclosed by the exit door and the vestibule door (this 
area will be relevant in our analysis).\\

\begin{figure}[!htbp]
\centering
\textbf{2-doors vestibule}\par\medskip
{\includegraphics[width=0.7\columnwidth]
{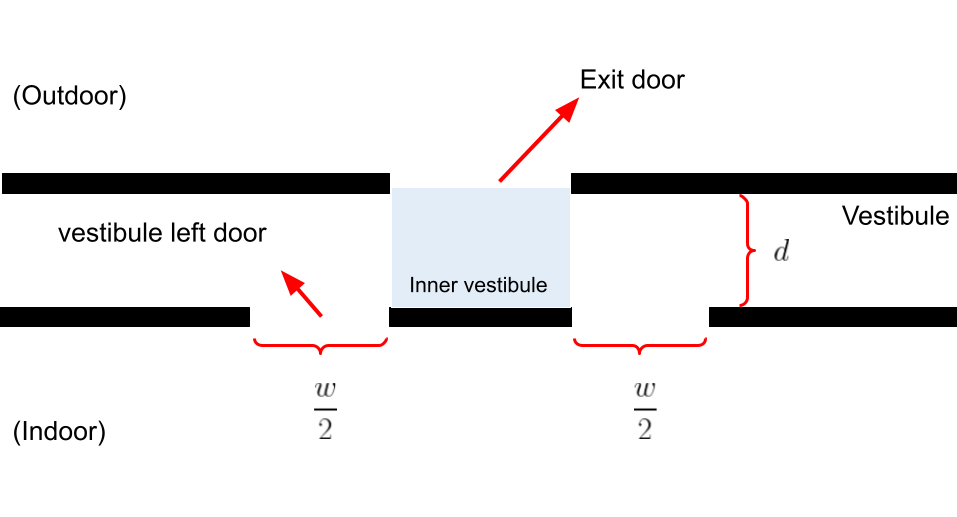}}\
\caption[width=0.49\columnwidth]{2-doors vestibule layout. $w/2$ is the width 
of each vestibule door. Both vestibule doors are symmetrically placed in front 
of the exit door. The vestibule is a corridor as long as the room and width 
$d$. The sides of the room are not drawn.}
\label{cv_2doors}
\end{figure}

The 2-doors vestibule layout is exhibited in Fig.~\ref{cv_2doors}. In this 
case, each vestibule door has a width $w/2$  ($w$ is the total vestibule door 
size). 
The two doors are separated by a wall similar in size to the exit door. This 
wall 
is located in front of the exit door.
The 2 vestibule doors are symmetrically located with respect to the exit. The 
shady area in Fig.~\ref{cv_2doors} represents the inner vestibule. In this case, 
it is 
defined as the rectangular area enclosed by the exit door and the opposing wall. 
\\

Notice that, we are dealing only with ``closed vestibules'' 
rather than ``open vestibules''. These are closed vestibules since the 
enclosing walls are as wide as the room. On the other hand, the open vestibules 
are characterized by one (or multiple) panel-like obstacles in front of the exit 
door. Previous research dealt with open vestibules where 
pedestrians were able to dodge the obstacles~\cite{sticco2022improving} or even 
step over them~\cite{ding2020evacuation} in order to access the vestibule. 
Instead, in the closed vestibule scenario, pedestrians have to go through the 
vestibule doors to access it.\\

\subsection{Clarifications}

We will refer to $d$ (the vestibule width) and $w$ (the 
vestibule door width) as the ``structural parameters''. In order to keep a 
clear notation, we will express the structural parameters ($d$ and $w$) in units 
of the agent's diameter. For example, if $w=4$, we actually mean 
that $w=4 \times 0.46\,$m $= 1.84\,$m.  Where 0.46\,m is the average 
pedestrian's shoulder-to-shoulder distance reported 
in~\cite{littlefield2008metric}.\\

Regarding the panel-like obstacles that define the vestibule, these panels
reach the sides of the room. Therefore, individuals can't dodge them. 
Moreover, in this research, panel-like obstacles are treated as walls (meaning
that individual-panel interactions are the same as the individual-wall 
interaction).\\

In this research, we use the agents' overlap as a reasonable indicator for the 
pressure exerted 
on them. We are aware that the overlap does not have units of pressure. 
Nevertheless, we argue that any reasonable measure of pressure should be 
correlated with the overlap since all the forces acting in the normal direction 
are monotonically non-decreasing functions of the overlap. In Section 
~\ref{pressure_subsec}, we will use the term ``overlap'' as a synonym of 
``pressure''. \\

Throughout this paper, we use the symbol $\rho$ to denote the 
density in the inner vestibule region. In the same way, we use the 
symbol $J$ to refer to the evacuation flow.\\

\section{\label{results} Results and discussions}

This Section is divided into three parts. In the first part 
(Section \ref{1-door}), we show the 
results corresponding to the 1-door vestibule layout for N=200. The second part 
corresponds to the 2-doors vestibule layout (Section \ref{2-doors}). These two 
sub-sections focus on the 
evacuation flow improvement that is possible to achieve with the aforementioned 
layouts. The explanation for this improvement is based on the 
fundamental diagram and the blocking cluster probability (also discussed in the 
aforementioned Sections).\\

Finally, in Section \ref{pressure_subsec} we further examine the effects of both 
vestibule layouts on the crowd pressure. We compare one layout against the other 
and also against the no-vestibule condition.

\subsection{\label{1-door}The 1-door vestibule}

We present in this Section the main results corresponding to the 1-door 
vestibule layout that was previously introduced in Fig.~\ref{cv_1door}. We 
explore a wide range of values for the structural parameters $d$ and $w$ and its 
consequences on the evacuation flow. Under some structural conditions, the presence 
of a 1-door vestibule improves the evacuation. At the end of the 
Section, we provide an explanation of the flow improvement given by 
the relation between the density inside the vestibule and the blocking clusters.\\

\begin{figure}[!htbp]
\centering
\subfloat[]{\includegraphics[width=0.49\columnwidth]
{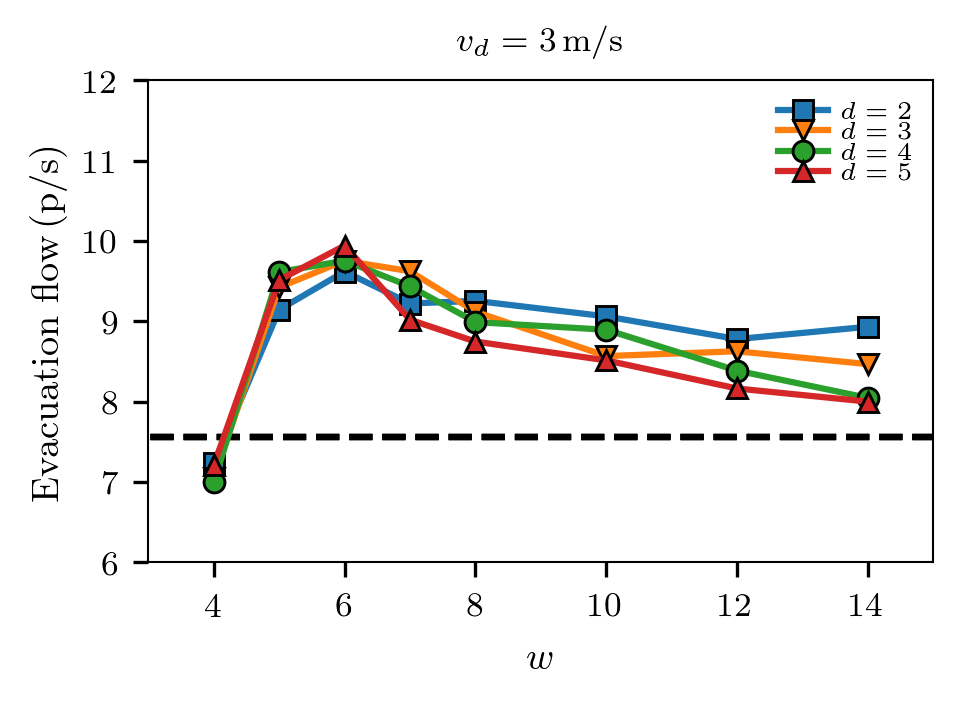}\label{flow_vs_w_alld_vd3_1door}}\ 
\subfloat[]{\includegraphics[width=0.49\columnwidth]
{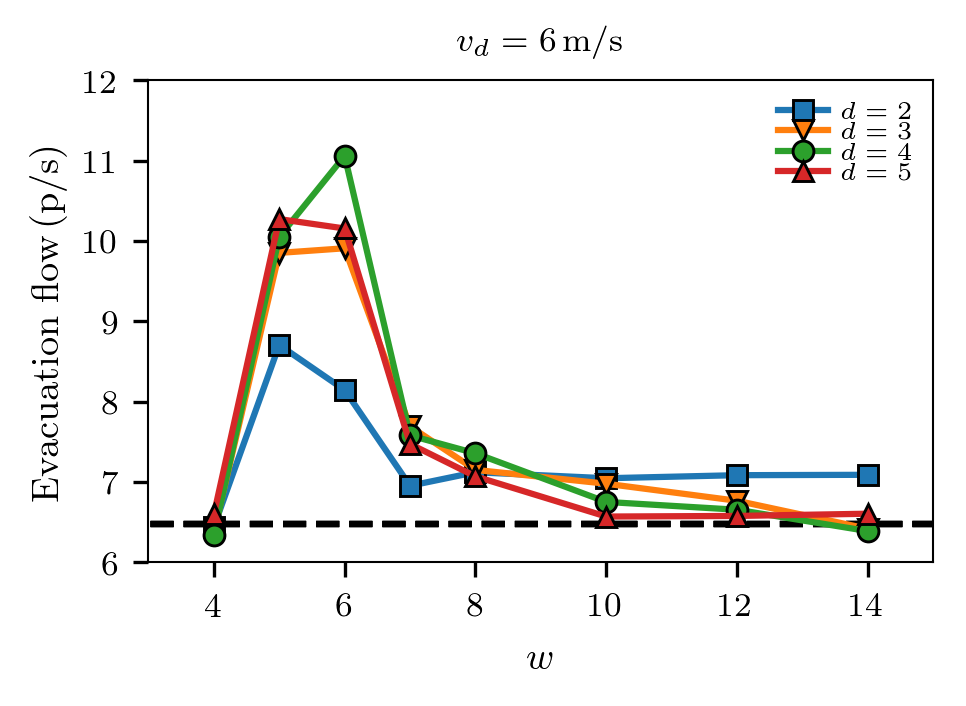}\label{flow_vs_w_alld_vd6_1door}}\\
\caption[width=0.47\columnwidth]{Evacuation flow as a function of the vestibule 
door width $w$ for 1-door vestibules. Each curve corresponds to a different 
vestibule width $d$. The horizontal dashed line indicates the flow for the 
``no-vestibule'' scenario. Data was averaged over 30 evacuation processes with 
random initial positions and velocities. The initial number of 
pedestrians was N=200. \textbf{(a)} the desired velocity was $v_d = 3\,$m/s, 
\textbf{(b)} the desired velocity was $v_d = 6\,$m/s.}
\label{flow_vs_w_1door}
\end{figure}

Fig.~\ref{flow_vs_w_1door}  shows the evacuation flow as a function of the 
vestibule door size ($w$) for different vestibule widths ($d$). The 
evacuation flow is defined in Eq.~(\ref{J}). The horizontal dashed line in 
Fig.~\ref{flow_vs_w_1door} stands for the ``no-vestibule'' situation. If a curve 
surpasses the horizontal dashed line, it means that the 
vestibule yields an enhanced evacuation flow with respect to the no-vestibule 
situation. Each plot corresponds to a different desired velocity $v_d$ as  
indicated in the plot's titles.\\

If $w$ is small ($w \leq 4$), the evacuation flow of the vestibule situation 
is equal to 
(or less than) the no-vestibule situation. On the opposite limit ($w>10$),
the evacuation flow also converges to the no-vestibule situation. 
Notably, intermediate $w$ values yield an evacuation flow that is much higher 
than the no-vestibule situation. This behavior holds for any $v_d$ in the 
range 2.5\,m/s$\leq v_d \leq 6\,$m/s. Fig.~\ref{flow_vs_w_alld_vd3_1door} and 
Fig.~\ref{flow_vs_w_alld_vd6_1door} show the results corresponding to 
$v_d=3\,$m/s and $v_d=6\,$m/s, respectively.\\

Notice that, regardless of $v_d$, varying $d$ does not 
affect the qualitative behavior of the flow curves. However, the evacuation 
flow strongly depends on $w$. Therefore, in the 1-door vestibule, we can assure 
that the evacuation flow is dominated by the structural parameter $w$ rather 
than $d$.\\

The largest difference between the 1-door vestibule situation and the 
no-vestibule scenario is given in $d=4$, $w=6$ and $v_d=6\,$m/s. In this case, 
the flow increment is more than 4 p/s (which represents a 70 \% increment). 
In a real-life environment, such a big contrast might be the difference 
between a regular evacuation and a disastrous event.\\

At a first inspection, the flow increment provided by the vestibule can 
be explained by the density in the ``inner vestibule''.\\

Recall that, the fundamental diagram (flow-density plot) is a meaningful chart 
for the ``free-flow'' regime (where the flow increases as the density 
increases) and the ``congested'' regime (where the flow diminishes as the 
density increases). The former is associated with a low-density scenario, 
whereas the latter is associated with a high-density scenario.\\

\begin{figure}[!htbp]
\centering
\subfloat[]{\includegraphics[width=0.49\columnwidth]
{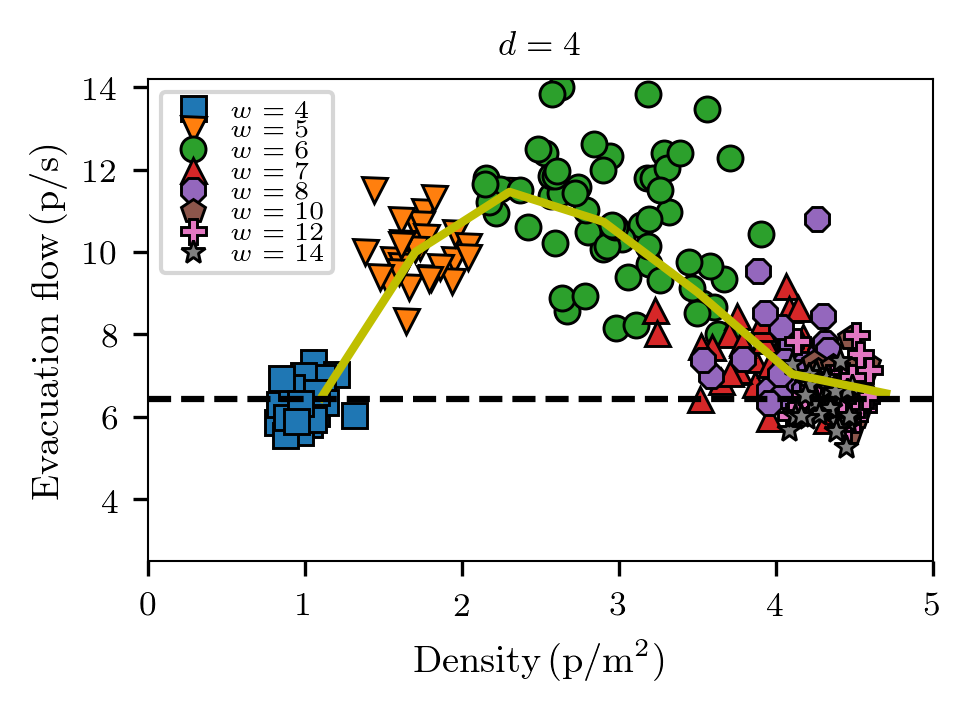}\label{flow_vs_density_vd6_cv_d4_1door}}
\ 
\subfloat[]{\includegraphics[width=0.49\columnwidth]
{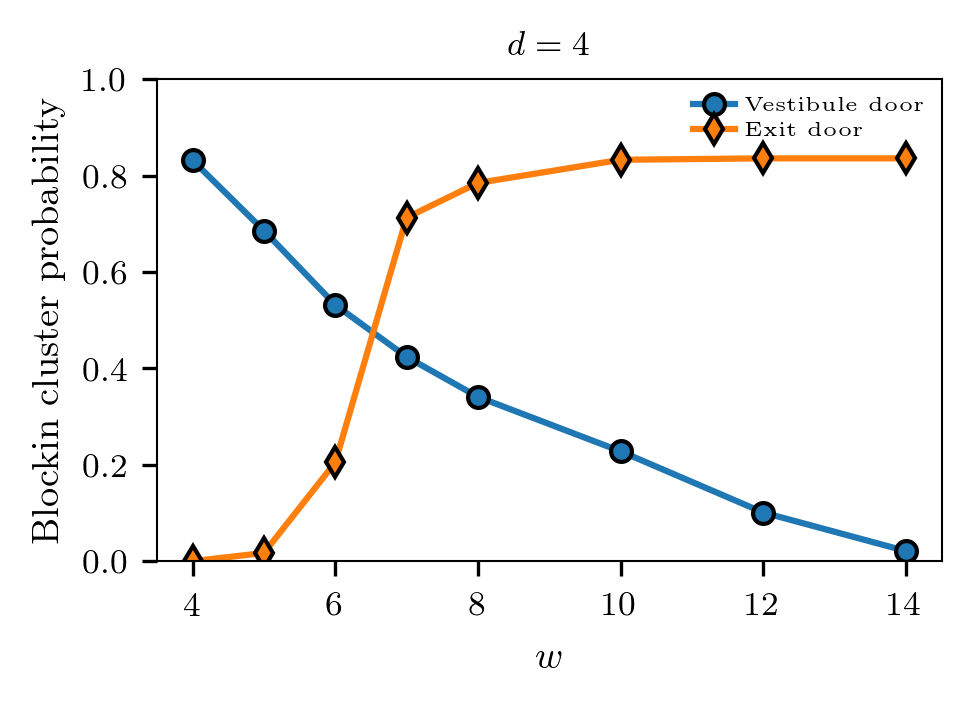}\label{pbc_vs_w_vd6_d4_cv}}\\
\caption[width=0.47\columnwidth]{\textbf{(a)} Evacuation flow as a function of 
the density (fundamental diagram) for 1-door vestibules. The horizontal dashed 
line indicates the flow for the ``no-vestibule'' scenario.
The density was measured on the inner vestibule and, it was averaged until 90\%
of individuals evacuated. Each data point
belongs to a single evacuation process where the initial condition 
(positions and velocities) were set to random.  \textbf{(b)} Blocking cluster 
probability as a function of the vestibule door width $w$. 
The probability is defined as the fraction of time that a blocking cluster  is 
present over the total evacuation time. The initial number of pedestrians was N= 
200. The desired velocity was $v_d = 6\,$ m/s.}
\label{flow_vs_density_1door}
 \end{figure}

Fig.~\ref{flow_vs_density_vd6_cv_d4_1door} 
exhibits the fundamental diagram for the 1-door vestibule at $d=4$ and 
$v_d=6\,$m/s. 
The horizontal dashed line corresponds to the no-vestibule situation. The 
markers stand for the $w$ values and the solid curve is the 
average over the data points. The data points correspond to different initial 
conditions. \\

The lowest $w$ values yield a free-flow regime, while the largest $w$ 
produce a congested regime. Notice that $w=6$ is the door width for 
intermediate densities ($\rho\sim2.5\,$p/m$^2$) and the maximum evacuation flow. 
We will explain the details of this phenomenon at the end of this Section.\\

If the density on the inner vestibule is low (say, $\rho \sim 1\,$p/m$^2$), the 
evacuation flow is below the optimal condition because there is unused 
space left close to the exit door. This phenomenon makes the crowd to evacuate 
``in dribs and drabs''.\\

If the density is high ($\rho \sim 4\,$p/m$^2$), the evacuation flow is also 
suboptimal because the area close to the exit door gets congested and produces 
blocking clusters that hinder the flow.\\

However, the intermediate density values ($\rho \sim 2.5\,$p/m$^2$) trade-off 
the two above mentioned phenomenons and therefore maximize the evacuation 
flow.\\

To complete the picture, we calculated the blocking cluster probability at the 
exit door and also at the vestibule door. A blocking cluster is defined as the 
set of pedestrians in physical contact that clog a door. Previous studies have 
shown that blocking clusters play a critical role in high-anxiety evacuation 
processes~\cite{parisi2005microscopic,cornesmicroscopic}. \\

Fig.~\ref{pbc_vs_w_vd6_d4_cv} shows the blocking cluster probability as a 
function of the 
vestibule door size $w$. Recall that, the blocking cluster probability is 
defined as the fraction of time that a blocking cluster is present over the 
total evacuation time.\\

The vestibule door curve and the exit door curve display an 
anti-correlation pattern (see Fig.~\ref{pbc_vs_w_vd6_d4_cv}). If the vestibule 
door is small ($w\leq 6$), there is a high blocking cluster probability at the 
vestibule door. The blocking clusters at the vestibule door prevent 
the inner vestibule from having high densities. This is why, for small $w$, the 
presence of blocking clusters at the exit door is almost negligible ($<0.2$).\\

Increasing $w$ reduces the blocking clusters at the vestibule door, letting a 
larger number of pedestrians access the inner vestibule at the same time. As a 
consequence, the exit door blocking cluster probability exhibits a sharp 
increase at $w=7$. Notice that, this increment is in agreement with the flow 
reduction of the congested regime exhibited at the  fundamental diagram 
(see Fig.~\ref{flow_vs_density_vd6_cv_d4_1door}).\\

It is worth clarifying that, although we only show the fundamental diagram 
and the blocking clusters corresponding to $d=4$, the qualitative behavior at 
different $d$ values is similar. In the same way, despite we only exhibit 
results for $v_d=3\,$m/s and $v_d=6\,$m/s, the qualitative behavior of the 
results for any desired velocity in the interval $v_d\geq2.5\,$m/s is similar to 
the ones presented here.\\

To conclude the Section, we remark that the evacuation flow is sensible 
to the inner vestibule density. At the same time, this density 
can be controlled mainly by the structural parameter $w$. Our most important 
result is that the evacuation flow gets maximized when the inner vestibule 
achieves a maximum density value such that almost no blocking clusters are 
produced at the exit door but not as low as to evacuate ``in dribs and 
drabs''.\\

\subsection{\label{2-doors}The 2-doors vestibule}

In this Section, we present the results corresponding to the 2-doors 
vestibule layout (illustrated in Fig.~\ref{cv_2doors}). This Section is 
organized in a similar fashion to the previous 
one. First, we explore the structural parameters ($d$ and $w$) that improve the 
evacuation flow and its relation to the fundamental diagram.  Then, we 
discuss about the blocking 
clusters' role in relation to the fundamental diagram and the evacuation 
performance. \\ 

\begin{figure}[!htbp]
\centering
\subfloat[]{\includegraphics[width=0.49\columnwidth]
{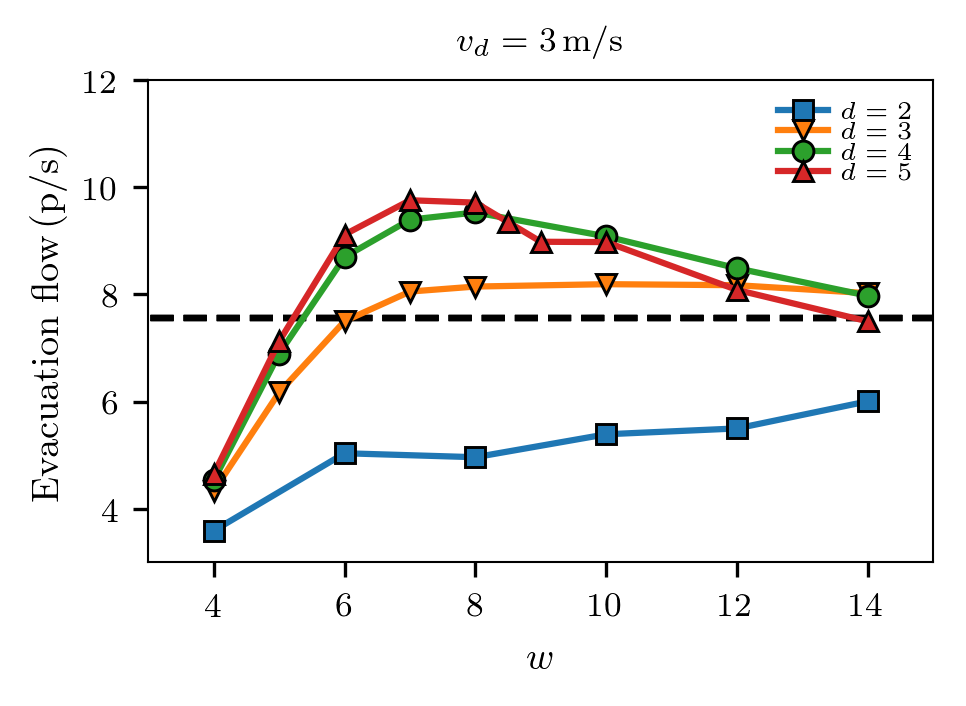}\label{flow_vs_w_alld_vd3_2doors}}\ 
\subfloat[]{\includegraphics[width=0.49\columnwidth]
{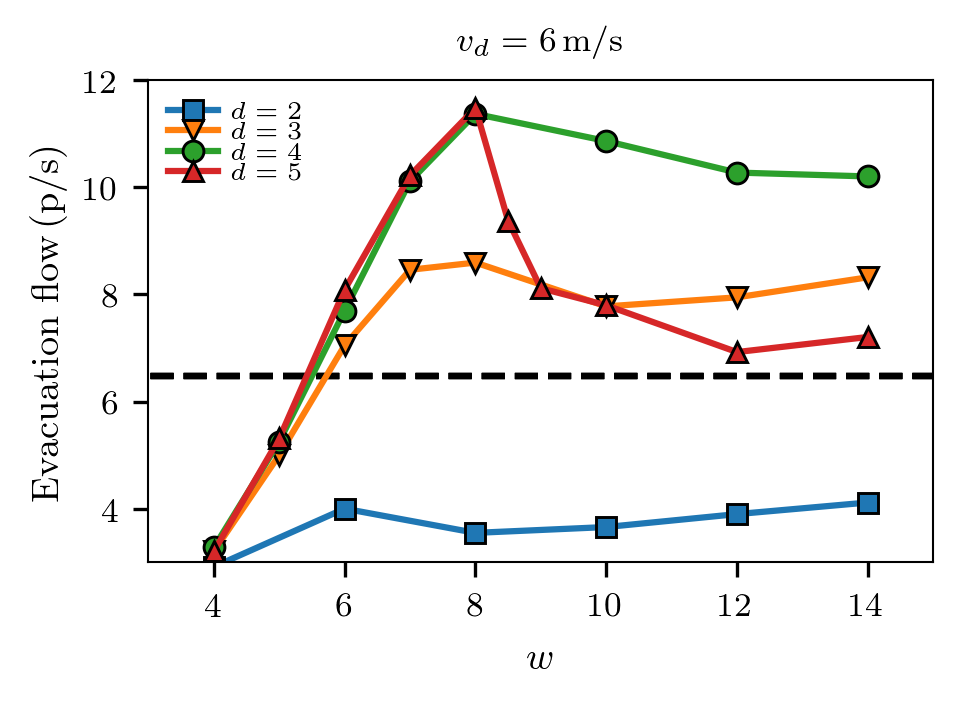}\label{flow_vs_w_alld_vd6_2doors}}\\
\caption[width=0.47\columnwidth]{Evacuation flow as a function of the total 
vestibule door width $w$ for 2-doors vestibules. Each curve corresponds to a 
different vestibule width $d$. The horizontal dashed line indicates the flow for 
the ``no-vestibule'' scenario. Data was averaged over 30 evacuation processes 
where the initial positions and velocities were set to random. The initial 
number of 
pedestrians was N=200. \textbf{(a)} the desired velocity was $v_d = 3\,$m/s, 
\textbf{(b)} the desired velocity was $v_d = 6\,$m/s.}
\label{flow_vs_w_2doors}
\end{figure}

The evacuation flow as a function of the vestibule's total door size ($w$) is 
shown in Fig.~\ref{flow_vs_w_2doors}. Since we are dealing with 2-doors 
vestibules, $w$ now stands for 
the sum of both vestibule door sizes. Recall that in this paper we only explore 
symmetrical vestibules. That is, the size of each door is $w/2$. \\

Fig.~\ref{flow_vs_w_alld_vd3_2doors} corresponds to $v_d = 3\,$m/s, while 
Fig.~\ref{flow_vs_w_alld_vd6_2doors} corresponds to $v_d = 6\,$m/s. Each curve 
stands for a different $d$ value (see legends for details). The 
no-vestibule evacuation flow is shown by 
the horizontal dashed line. As a first inspection, we can notice that $d=2$ 
worsens the evacuation performance, but $d=3$ increases the evacuation flow 
until reaching a plateau above the no-vestibule situation. \\

In the case of $v_d=3\,$m/s (Fig.~\ref{flow_vs_w_alld_vd3_2doors}), the widest 
vestibules ($d=4$ and $d=5$) exhibit 
similar patterns. Small $w$ values worsen the evacuation performance but, the 
evacuation flow improves for $w\geq6$. In both cases, the curves reach a maximum 
at $w=8$ and then reduce the flow as $w$ increases.\\

In the case of $v_d=6\,$m/s (Fig.~\ref{flow_vs_w_alld_vd6_2doors}), the 
vestibule of size $d=4$ produces the maximum 
flow at $w=8$ (the reader may watch the video in the supplementary material). 
After this ``peak'', the flow slowly diminishes for increasing 
values of $w$. It is worth noting that, for $w>6$, the evacuation flow is 
substantially higher than the no-vestibule situation.\\

If the vestibule is wider ($d=5$) and $v_d=6\,$m/s (see 
Fig.~\ref{flow_vs_w_alld_vd6_2doors}), the curve also exhibits an 
interval in which the evacuation flow is much higher than the no-vestibule 
situation. The ``peak'' of flow is around $w=8$ and the flow converges to the 
no-vestibule scenario as $w$ increases. \\

It is important to remark that the 2-doors vestibule is capable of improving the 
evacuation flow under a wide range of values of the structural parameters 
($d$ and $w$).\\

Interestingly, the maximum flow value for 2-doors vestibule is around $w=8$, 
whereas the maximum flow for the 1-door vestibule is around $w=6$. It means 
that a single door of size $w$ yields more flow than 2 doors of size $w/2$. 
Similar results were reported for two-doors pedestrian evacuation 
simulations~\cite{sticco2017room} and also for experiments performed with 
granular media accelerated by the force 
of gravity~\cite{fullard2019dynamics}. In ~\ref{appendix_1} we provide a brief 
discussion regarding this 
topic.\\ 

The vestibule improves the evacuation flow if certain structural conditions are 
met (specific values of $d$ and $w$). At a first inspection, the flow increment 
can be explained with the help of the fundamental diagram (\textit{i.e.} the 
flow vs. density plot). As we did for the 1-door vestibule, we measured the 
evacuation flow as a function of the density. The density was measured in the 
``inner vestibule'' (the area between the exit door and the vestibule wall; 
see Section \ref{layouts} for details). \\

\begin{figure}[!htbp]
\centering
\subfloat[]{\includegraphics[width=0.48\columnwidth]
{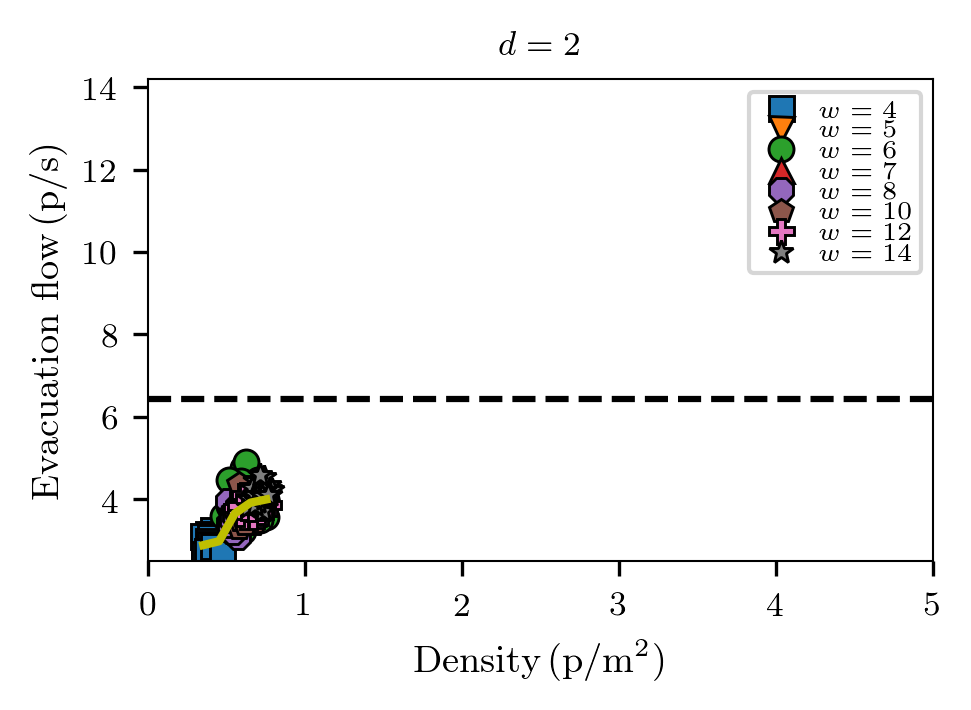}\label{flow_vs_density_vd6_cv_d2_2doors}
}\ 
\subfloat[]{\includegraphics[width=0.48\columnwidth]
{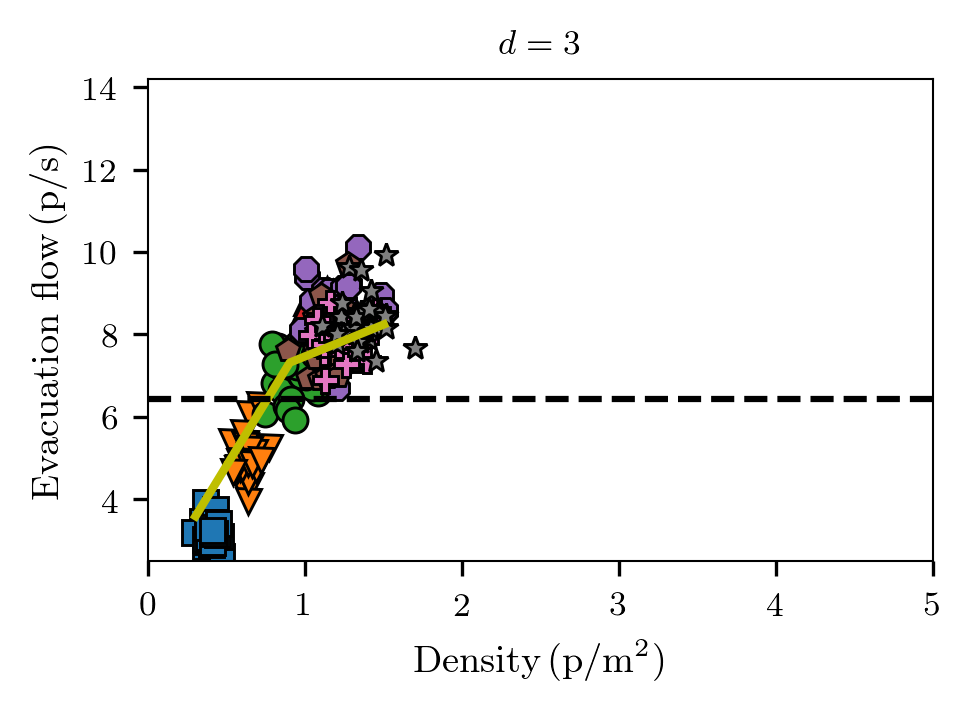}\label{flow_vs_density_vd6_cv_d3_2doors}
}\\
\subfloat[]{\includegraphics[width=0.48\columnwidth]
{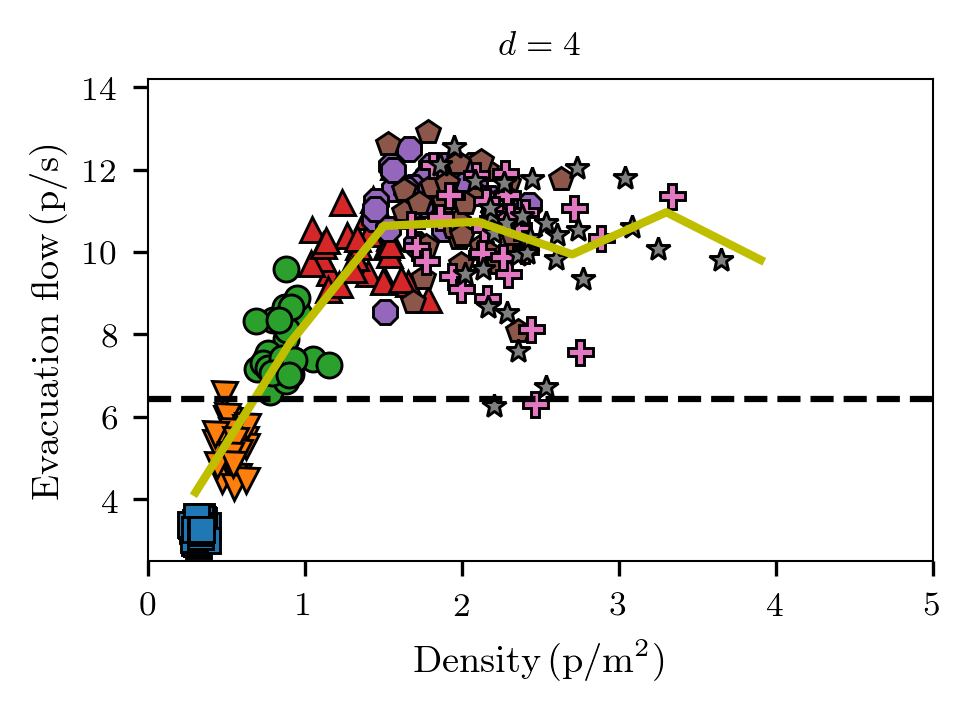}\label{flow_vs_density_vd6_cv_d4_2doors}
}\ 
\subfloat[]{\includegraphics[width=0.48\columnwidth]
{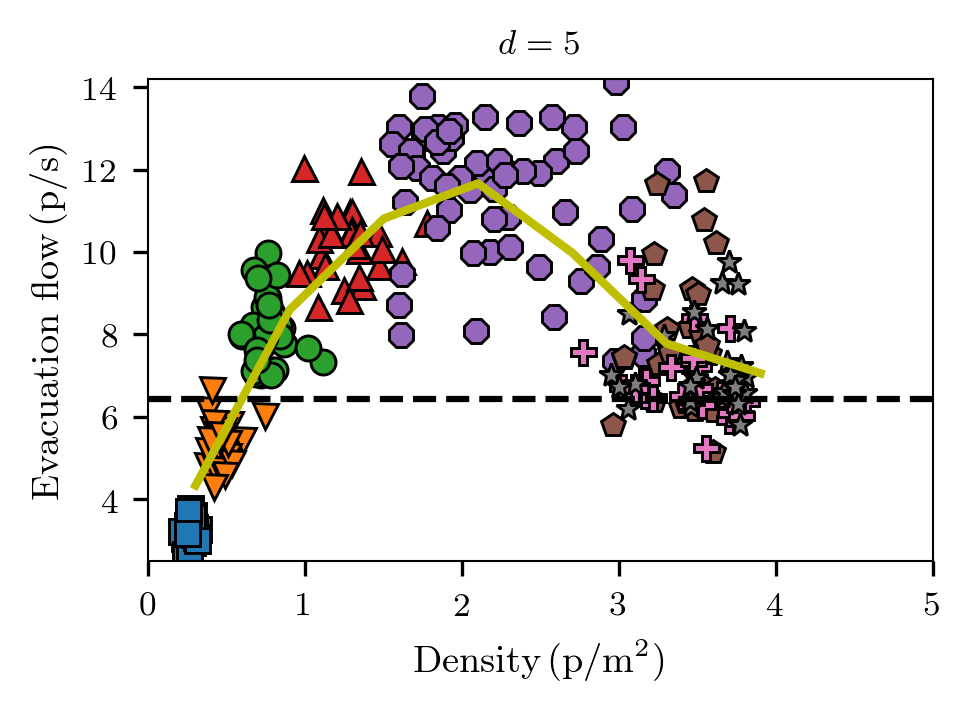}\label{flow_vs_density_vd6_cv_d5_2doors}
}\\
\caption[width=0.47\columnwidth]{Evacuation flow as a function of the density 
(fundamental diagram) for 2-doors vestibules. The density was measured on the 
inner vestibule and, it was averaged until 90\% of individuals evacuated. The 
horizontal dashed line indicates the flow for the ``no-vestibule'' scenario. 
Each data point belongs to a single evacuation process where the initial 
condition (positions and velocities) were set to random. The initial number of 
pedestrians was N=200. The desired velocity was $v_d = 6\,$m/s. See the plot's 
title for the corresponding value of the vestibule width $d$.}
\label{flow_vs_density_2doors}
\end{figure}

Fig.~\ref{flow_vs_density_2doors} shows the fundamental diagram for different 
structural parameters 
(see the titles and the labels). The horizontal line stands for the 
no-vestibule situation. The markers represent different vestibule door size 
$w$ and each data point correspond to a different initial condition. \\

The narrowest vestibule explored is $d=2$. Under this condition, the density in 
the inner vestibule remains very low ($\rho < 1$) regardless of the size of 
the vestibule door (Fig.~\ref{flow_vs_density_vd6_cv_d2_2doors}). The cause of 
this phenomenon is that pedestrians get stuck 
at the entrance of the inner vestibule preventing this zone to have a larger 
number of pedestrians. The direct consequence of this phenomenon is a lower 
evacuation flow than that of the situation without a vestibule.\\

Fig.~\ref{flow_vs_density_vd6_cv_d3_2doors} corresponds to 2-doors vestibule and 
$d=3$. Although this structural 
condition allows a wider range of density and flow, only the free-flow regime is 
observed. This means that the inner vestibule does not get overcrowded 
(regardless of $w$). This phenomenon is related to the blocking clusters produced
inside the vestibule, as will be discussed at the end of this 
Section.\\

For 2-doors vestibules of size $d=4$ 
(Fig.~\ref{flow_vs_density_vd6_cv_d4_2doors}), the free-flow and the 
``peak'' of the fundamental diagram are observed. This means that increasing 
$w$ increases the density while also increasing the evacuation flow. 
As in the case of $d=3$, the inner vestibule does not get overcrowded because the
agents get stuck (temporarily) before the inner vestibule region. \\

If the vestibule is wider ($d=5$), it is possible to observe the free-flow and 
the congested regime (Fig.~\ref{flow_vs_density_vd6_cv_d5_2doors}). The former 
holds for $w<8$, while the latter appears at $w>9$. The congested regime is 
characterized by decreasing flow as the density increases. For such a wide 
vestibule ($d=5$), and such a large vestibule door size ($w>8$), a large number 
of agents are allowed to enter the vestibule at the same time. This is the limit 
that resembles the no-vestibule scenario. \\

To complete the picture of why the inner vestibule gets crowded depending on
$d$ and $w$, we measured the blocking clusters. Unlike the 1-door vestibule,
for the 2-doors vestibule, it is necessary to define a new type of blocking cluster:
the ``inside vestibule blocking cluster''. These are blocking clusters that are 
formed inside the vestibule but just before the inner vestibule. \\

In summary, three types of blocking clusters play a relevant role. The ones that are 
formed at the exit door, the ones that are formed before the vestibule (at the 
vestibule door), and the ones that are formed inside the vestibule (just before 
the inner vestibule). See \ref{appendix_bc} for an illustration of 
the blocking cluster types. In any case, the blocking cluster probability is 
defined as the fraction of time that blocking clusters are present over the 
total evacuation time.\\

\begin{figure}[!htbp]
\centering
\subfloat[]{\includegraphics[width=0.49\columnwidth]
{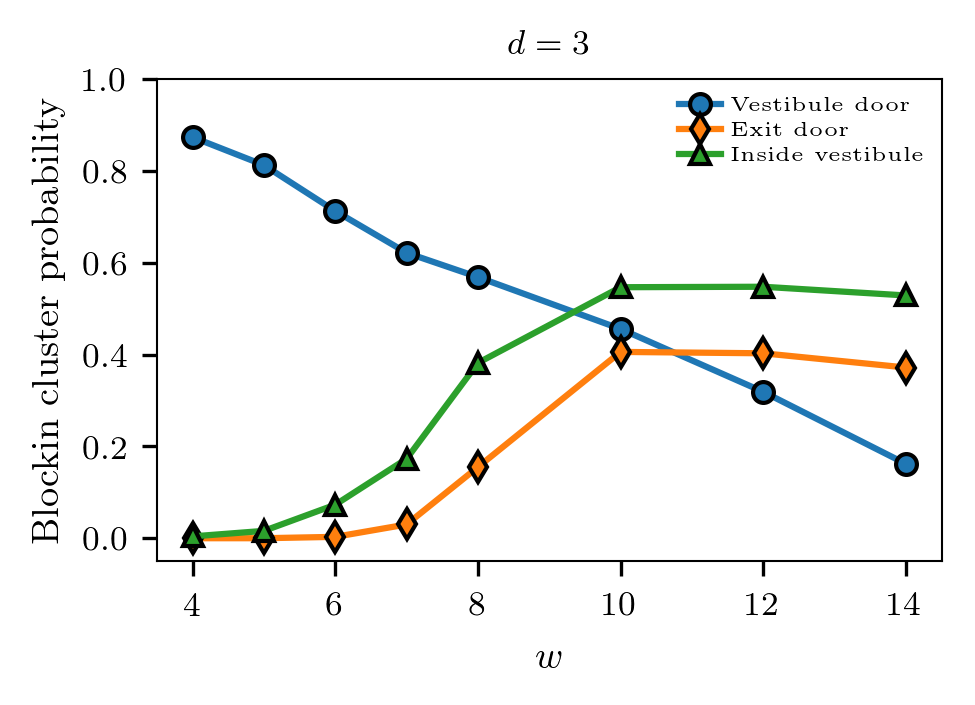}\label{pbc_vs_w_vd6_d3_cv_2doors}}\ 
\subfloat[]{\includegraphics[width=0.49\columnwidth]
{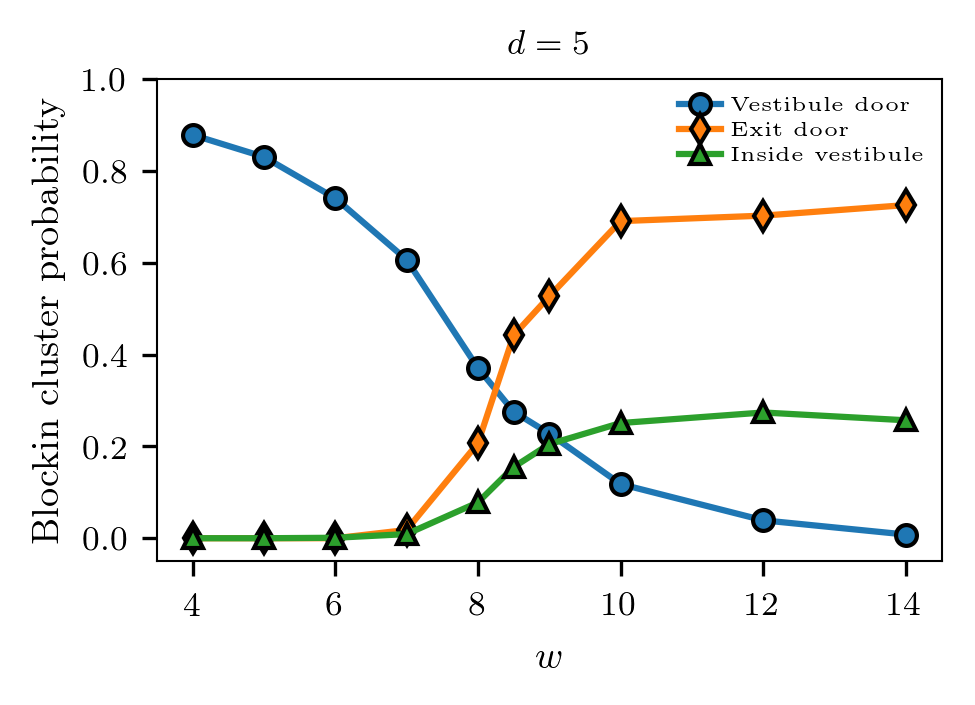}\label{pbc_vs_w_vd6_d5_cv_2doors}}\\
\caption[width=0.47\columnwidth]{Blocking cluster probability as a function of 
the vestibule door width $w$. See the legend for the type of considered blocking 
cluster. The probability is defined as the fraction of time that a blocking 
cluster is present over the total evacuation time. 
The initial number of pedestrians was N=200. The desired velocity was
$v_d = 6\,$m/s. \textbf{(a)} For vestibule width $d=3$, \textbf{(b)} For 
vestibule width $d=5$.}
\label{pbc_2doors}
\end{figure}

Fig.~\ref{pbc_vs_w_vd6_d3_cv_2doors} and Fig.~\ref{pbc_vs_w_vd6_d5_cv_2doors} 
show the three types of blocking cluster probabilities for 
$d=3$ and $d=5$, respectively. As expected, increasing $w$ reduces the 
vestibule door blocking clusters. The main difference between the case for $d=3$ 
and $d=5$ is that, in the former, the ``inside vestibule curve''  is above the 
``exit door curve''. On the other hand, for $d=5$, the ``exit door curve'' is 
well above the ``inside vestibule curve''. \\

The blocking cluster curves explain the differences in the fundamental diagram. 
$d=3$ increases the blocking cluster probability before the inner vestibule.
These blocking clusters prevent the inner vestibule 
from getting overcrowded. This is why the fundamental diagram at $d=3$ only 
exhibits free-flow (and low density). The direct consequence of the low density 
is the lack of blocking clusters at the exit door. Therefore, the evacuation 
flow is higher than the no-vestibule condition. \\

The scenario of $d=5$ produces fewer blocking clusters before the inner 
vestibule. Thus, the inner vestibule gets crowded more easily. This is why 
the fundamental diagram at $d=5$ exhibits a congested regime (for large enough 
$w$). The consequence is an increase in the exit door blocking cluster 
probability, which reduces the evacuation flow. \\

To conclude this Section, we stress that the evacuation flow in the 2-doors 
vestibule is strongly dependent on the density. However, it is possible to 
control the inner vestibule density with the structural parameters $d$ and $w$. 
In the same way as the 1-door vestibule, the flow is maximized for intermediate 
density values ($\rho\sim2.5\,$p/m$^2$). Under this condition, the density is 
low enough to minimize the formation of exit door blocking clusters, but not as 
low as to leave unused space left in the vestibule.\\

\subsection{\label{pressure_subsec}The pressure}

In the previous Section, we showed that under certain structural conditions 
the inclusion of a vestibule at the exit increases the evacuation flow. One 
concern now is whether the improvement in flow is at the expense of an increase 
in the pressure suffered by the escaping individuals. 
This possibility arises because the vestibule is achieved by placing 
panel-like obstacles and, these obstacles may cause an increase in pressure due 
to the individual-obstacle interaction. Another arguable matter is whether the 
vestibule may create higher local density regions where the pressure is 
maximized. \\

In this Section, we report the ``overlap'' between agents as an indicative of the
pressure. The overlap on the $i-$th particle is defined as $o_i = \sum_{j} 
\left [ R_{ij}-d_{ij} 
\right ]$ where $R_{ij}=R_i+R_j$ is the sum of radius of particles $i$ and $j$. 
$d_{ij}$ is the distance between mass centers. $j$ stands for any particle (or 
wall) that is in physical contact with particle $i$. Thus, the overlap reflects 
the degree of closeness between pedestrians. We stress that in this Section, we 
use the terms overlap and pressure interchangeably.\\

Although the overlap is not the unique way of quantifying the pressure in a 
crowd~\cite{feliciani2020systematic,helbing2007dynamics, 
garcimartin2017pedestrian}, we argue that any reasonable measure of 
pressure should correlate with this magnitude. We support this argument 
because the forces acting in the normal direction are monotonically 
non-decreasing functions of the overlap.\\

\begin{figure}[!htbp]
\centering
\subfloat[]{\includegraphics[width=0.48\columnwidth]
{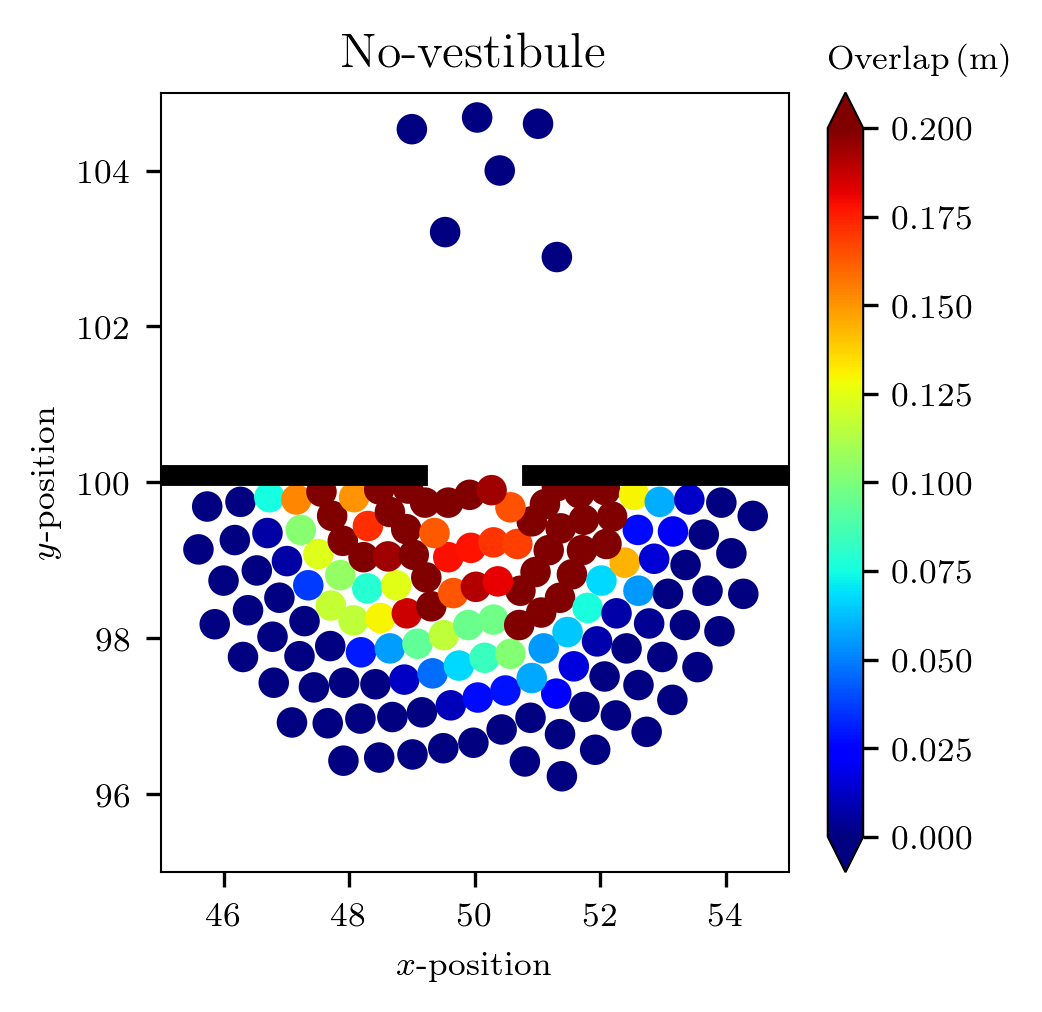}\label{overlap_novest_t7}
}\\ 
\subfloat[]{\includegraphics[width=0.48\columnwidth]
{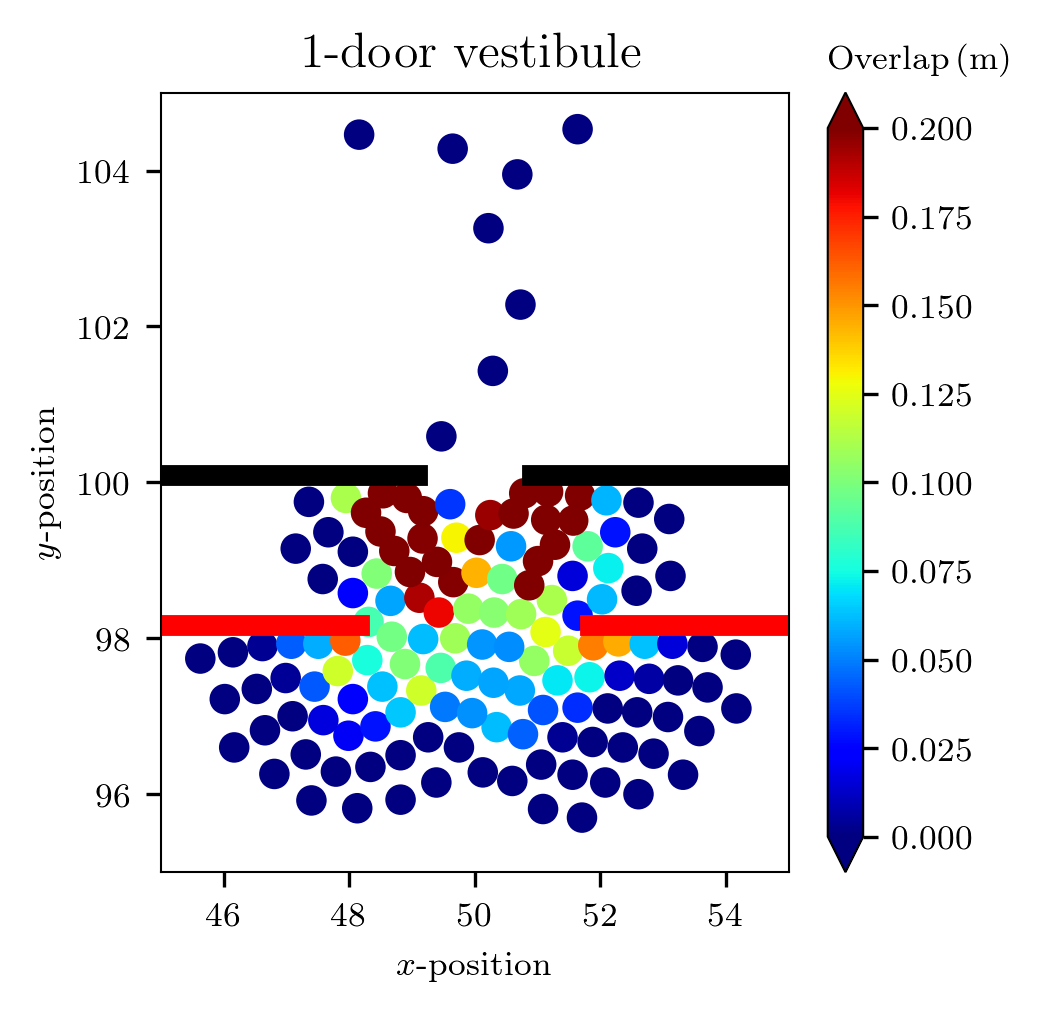}\label{overlap_1door_w4_t7}
}\
\subfloat[]{\includegraphics[width=0.48\columnwidth]
{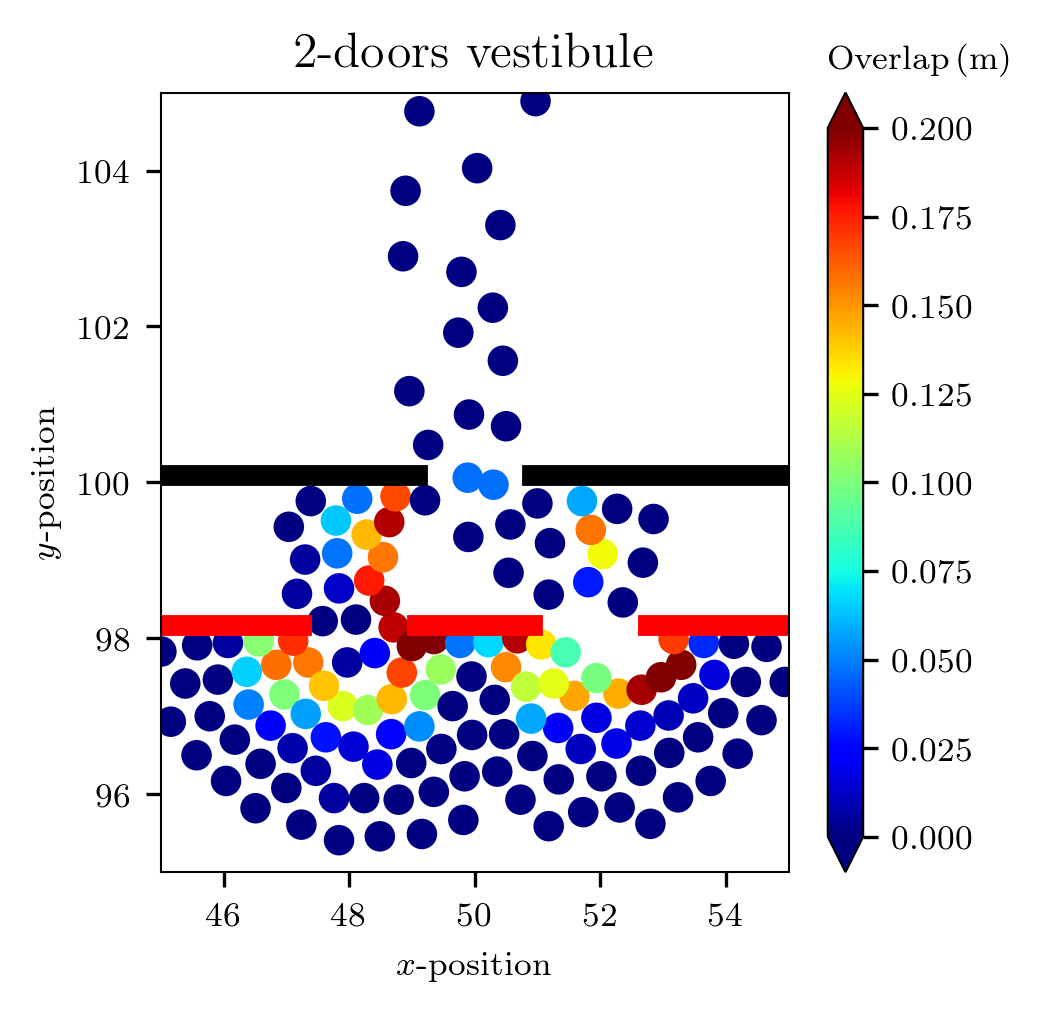}\label{overlap_2doors_w4_t7}
}\
\caption[width=0.47\columnwidth]{Snapshots of the numerical simulations at 
$t=7\,$s and $v_d=6\,$m/s. The circles represent the simulated pedestrians. The 
color on each agent stand for the overlap level (see scale on the right). The 
title on each plot refers to the layout condition. The 3 scenarios shown begin 
with the same initial conditions (positions and velocities) for N=200 
pedestrians. The vestibule door width is $w=$8 in both cases.}
\label{overlap_map}
\end{figure}

Our results show that the presence of a vestibule significantly reduces the 
pressure on pedestrians. As a first approach, we display 3 snapshots of 
evacuations for three different layouts: no-vestibule 
(Fig.~\ref{overlap_novest_t7}), 1-door vestibule 
(Fig.~\ref{overlap_1door_w4_t7}), and 2-doors vestibule 
(Fig.~\ref{overlap_2doors_w4_t7}). Each circle represents a simulated 
pedestrian. The color of each agent reflects the level of overlap $o_i$ 
(see caption for details).\\

It is easy to observe that the no-vestibule situation exhibits the highest 
pressure among the three situations explored. The 1-door vestibule produces 
lower pressure than the no-vestibule situation and, the 2-door vestibule 
scenario produces even less pressure than the 1-door vestibule 
(see Fig.\ref{overlap_map}). Two reasons explain the pressure reduction 
produced by the vestibule, as follows.\\


The first reason is that the interaction at the vestibule doors reduces the 
velocity of pedestrians inside the vestibule region (which reduces the pressure). 
In other words, the vestibule doors alleviate the strain on the escaping pedestrians at 
the exit. This phenomenon, explains the difference between the no-vestibule 
scenario and the vestibule condition.\\


The second reason that explains the pressure reduction concerns crowd 
dispersion. The more dispersed the crowd is, the lower the pressure. The 2-doors 
scenario shows a more dispersed crowd in the zone before the vestibule. In other 
words, the 2-door vestibule ``forces the crowd'' to split into two halves (one 
for each door) which produces an overall density reduction (hence a pressure 
reduction). The parameter $w$ is the key factor to control the crowd 
dispersion.\\

\begin{figure}[!htbp]
\centering
\subfloat[]{\includegraphics[width=0.49\columnwidth]
{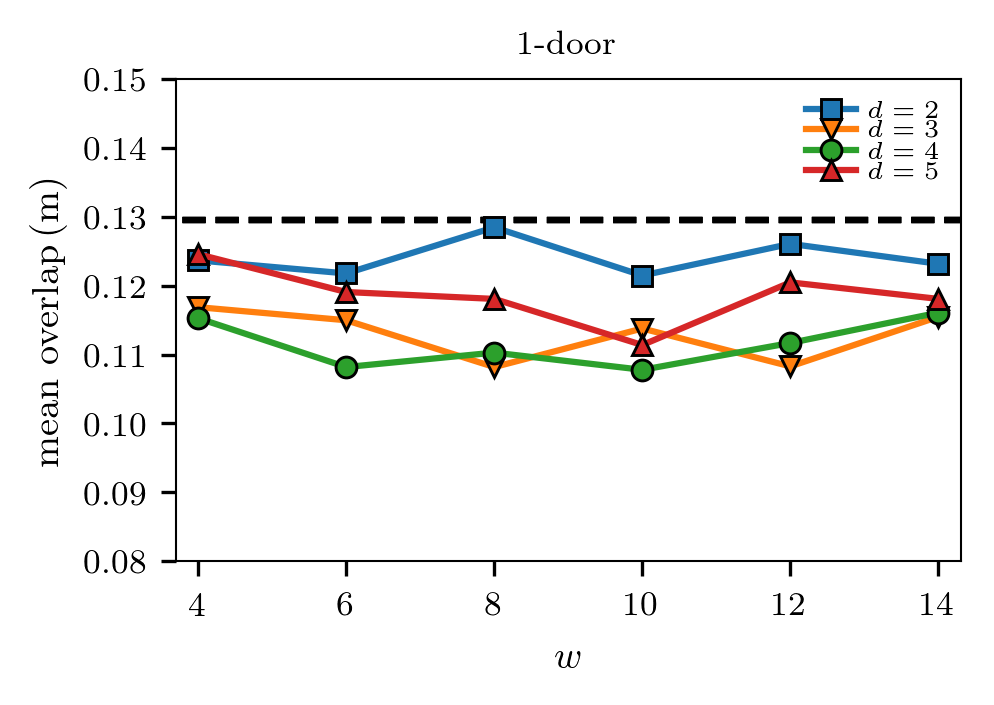}\label{overlap_vs_w_1door}}\ 
\subfloat[]{\includegraphics[width=0.49\columnwidth]
{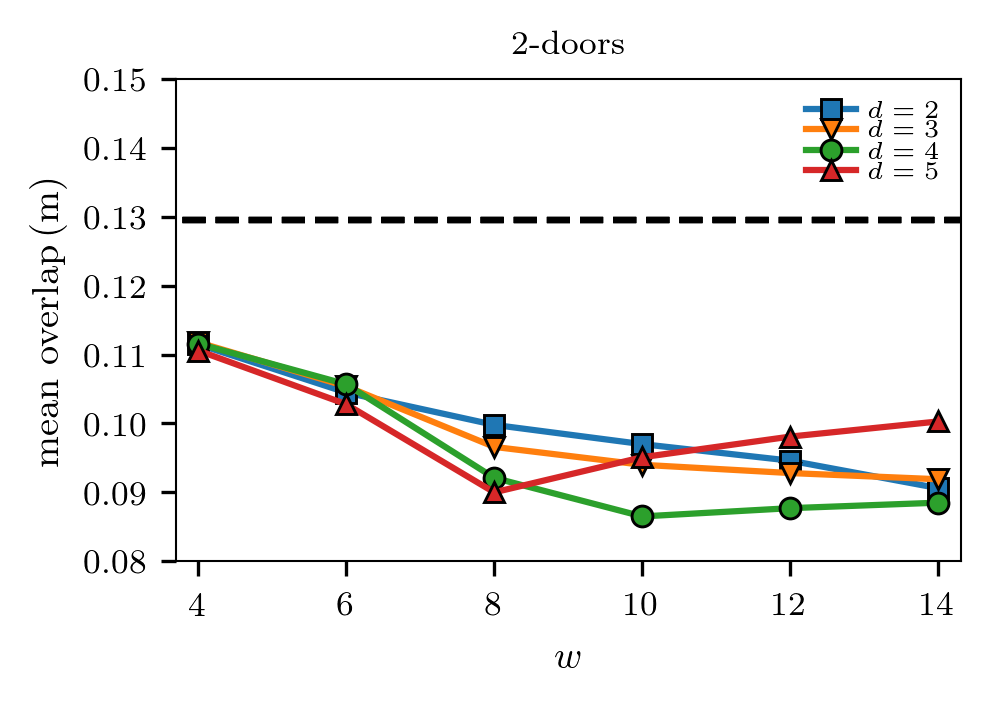}\label{overlap_vs_w_2doors}}\\
\caption[width=0.47\columnwidth]{Mean crowd overlap as a function of $w$ for 
different vestibule widths $d$. The horizontal dashed line represents the 
overlap for the no-vestibule situation. The overlap values are averaged over 
time for  30 different initial conditions. The simulations finished when 180 
agents left the room. The initial number of agents was N=200 and the desired 
velocity was $v_d=6\,$m/s.}
\label{overlap_vs_w}
\end{figure}

We computed the mean overlap to quantify the crowd pressure for 
different structural conditions ($d$ and $w$). This metric is the mean crowd 
overlap averaged over time at different initial conditions. 
Fig.~\ref{overlap_vs_w} shows the 
mean overlap as a function of $w$ for different vestibule widths $d$. 
Fig.~\ref{overlap_vs_w_1door} corresponds to the 1-door vestibule and 
Fig.~\ref{overlap_vs_w_2doors} corresponds to 
the 2-doors vestibule. The horizontal dashed line represents the mean overlap 
for the no-vestibule situation. \\

For all the explored conditions, the vestibule produces a lower mean overlap 
than the no-vestibule situation (all curves lies below the horizontal dashed 
line in Fig.~\ref{overlap_vs_w}). The 2-door vestibule scenario 
(Fig.~\ref{overlap_vs_w_2doors}) exhibits less pressure than the 1-door 
vestibule scenario (Fig.~\ref{overlap_vs_w_1door}) for any of the structural 
condition explored. Roughly, the 2-door vestibule shows a decreasing mean 
overlap as $w$ increases, whereas the 1-door vestibule presents a 
quasi-constant pattern for any $w$ value.\\

The decreasing trend in the 2-doors vestibule is a consequence of the crowd 
dispersion. Recall that, increasing $w$ means increasing the distance between 
the edges of the 2 doors (see Fig.\ref{cv_2doors}). Therefore, in the 2-doors 
scenario, the higher $w$, the more spread the crowd is. This phenomenon has a 
direct consequence on the pressure reduction.\\

On the other hand, the 1-door vestibule does not significantly spread out 
the crowd as $w$ increases because the crowd is not split into two halves 
(unlike in the 2-doors scenario). This is why the overlap exhibits a 
quasi-constant pattern in Fig.~\ref{overlap_vs_w_1door}.\\

Although the results presented in Fig.~\ref{overlap_vs_w} correspond to 
$v_d=6\,$m/s, we obtained similar patterns for desired velocities in the 
interval $v_d>2.5\,$m/s. We remark that, in addition to the 
pedestrian-pedestrian interaction, the pedestrian-wall interaction is taken into 
account in this analysis.\\


The above results can be summarized as follows. Our numerical simulations 
indicate that the presence of a vestibule reduces the 
overall pressure in the crowd. This result holds for any of the structural 
condition (varying $w$ and $d$). We highlight that the 2-door vestibule appears 
as the 
most effective way of reducing the crowd pressure. The key to this pressure 
reduction has two components. On one hand, the presence of the vestibule 
reduces the velocity inside the vestibule region which reduces the pressure.
On the other hand, the increase in the parameter $w$ allows the crowd to 
spread out, thus, reducing the density in the area next to the vestibule.\\


\section{\label{conclusions} Conclusions}

We propose architectural improvements to enhance the evacuation performance in 
emergency situations. We performed numerical simulations, in the context of the 
social force model, in order to recreate a crowd of N=200 pedestrians 
evacuating a room under a highly stressing situation ($v_d \geq 3\,$m/s). \\

We explored 3 layout conditions: the no vestibule scenario, the 1-door 
vestibule, and the 2-doors vestibule. The structural parameters that 
characterize the vestibule conditions are $d$ (the vestibule width) and $w$ (the 
vestibule door width). We found that an adequate selection of these parameters 
can substantially increase the evacuation flow. \\

The optimal evacuation flow ($J\sim 11\,$p/s) is achieved at $d=4$, $w=6$ for 
the 1-door vestibule and, at $d=4$, $w=8$ for the 2-doors vestibule. This is a 
remarkable improvement considering that the no-vestibule scenario yields 
$J\sim6.5\,$p/s (for $v_d=6\,$m/s in all cases).\\

The key to understanding this phenomenon is that $d$ and $w$ control the density 
in the inner vestibule region. Low density values ($\rho \sim 
1\,$p/m$^2$) produce a suboptimal evacuation flow because there is unused 
space left close to the exit door. This state makes the crowd to evacuate ``in 
dribs and drabs''.\\

Intermediate density values ($\rho \sim 2.5\,$p/m$^2$) maximize the evacuation 
flow because the inner vestibule receives the maximum possible amount of 
pedestrians without producing many blocking clusters at the exit door. In the 
opposite case, high density values ($\rho \sim 4\,$p/m$^2$), yield a suboptimal 
evacuation flow because it increases the probability of producing blocking 
clusters at the exit.\\ 

Another important achievement is the pressure reduction attained by the 
vestibule layouts (in comparison with the no-vestibule scenario). Although both 
of the explored vestibules (\textit{i.e.} the 1-door and the 2-doors vestibule) 
reduce the crowd pressure, it is the 2-doors vestibule the one that lowers the 
pressure the most. This phenomenon occurs because increasing $w$ in this layout 
forces the crowd to spread out more, thus reducing the local density in the 
area before the vestibule.\\

It is also quite relevant that this kind of architectural improvement do not 
require the pedestrians to be trained to adopt the expected behavior for 
a successful evacuation.\\

Although this is a numerical study, we believe that the principles elaborated 
here are strong enough to be tested empirically in future investigations. As a 
final word, we would like this paper to be a valuable source of inspiration for 
future research that truly seeks to improve the human condition.\\

\appendix

\section{\label{appendix_a}Evacuation flow for a larger crowd}

In order to make a brief assessment of the robustness of the results presented 
in this paper, we performed numerical simulations for a larger crowd. In this 
appendix, we show the flow vs. $w$ for different $d$ values for an initial crowd 
of size N=600 (remember that we have already shown results for N=200).\\

\begin{figure}[!htbp]
\centering
\subfloat[]{\includegraphics[width=0.49\columnwidth]
{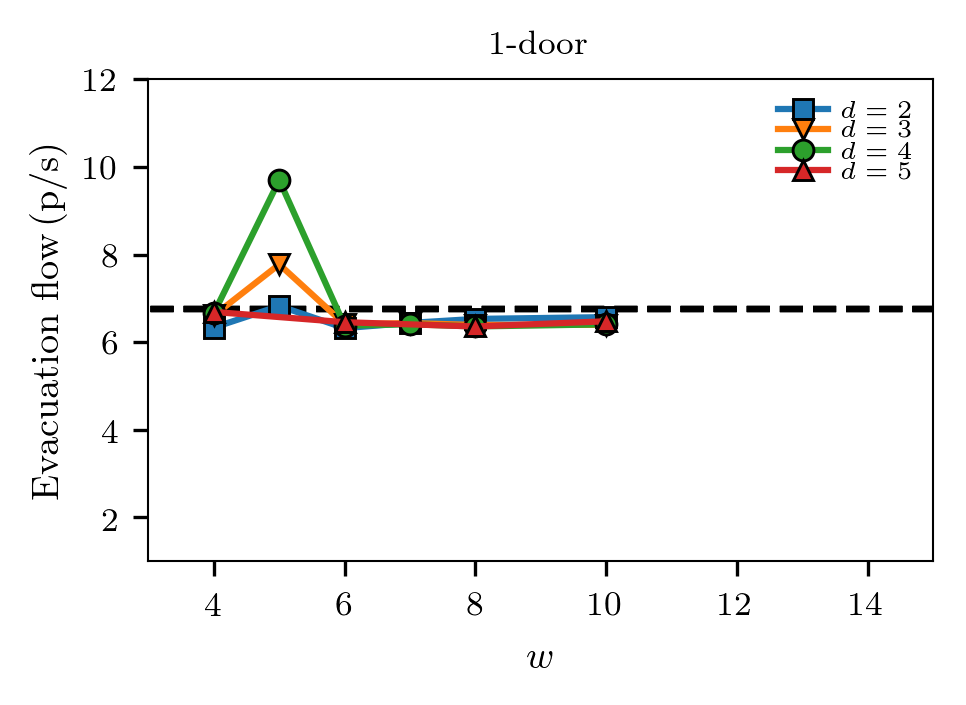}\label{flow_vs_w_N600_1door}}\ 
\subfloat[]{\includegraphics[width=0.49\columnwidth]
{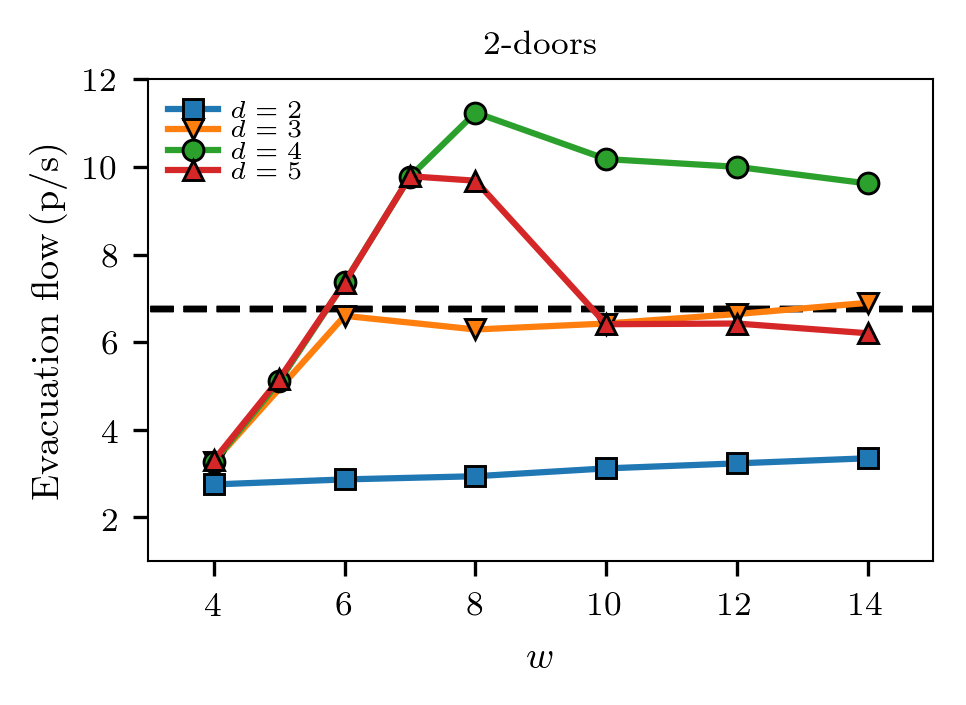}\label{flow_vs_w_N600_2doors}}\\
\caption[width=0.47\columnwidth]{Evacuation flow as a function of the vestibule 
door width $w$. Each curve corresponds to a different vestibule width $d$. The 
horizontal dashed line indicates the flow for the ``no-vestibule'' scenario. 
Data was averaged over 30 evacuation processes where the initial positions and 
velocities were random. The initial number of pedestrians was N=600. See the 
plot's title for the corresponding layout.}
\label{N600}
\end{figure}

Fig.~\ref{flow_vs_w_N600_1door} displays the results 
corresponding to the 1-door vestibule and Fig.~\ref{flow_vs_w_N600_2doors} shows 
the results corresponding to the 2-doors vestibule. The horizontal dashed line 
stands for the evacuation flow in the ``no-vestibule'' scenario at N=600.  The 
most remarkable characteristic is that the vestibule still proves to achieve a 
considerable flow increment for a crowd as large as N=600.\\

Both vestibule layouts seem to improve the evacuation performance. However, the 
2-doors vestibule yields higher flow values than the 1-door vestibule. Moreover, 
in the 2-door vestibule scenario, there is a wider range of structural 
parameters $d$ and $w$ that surpass the flow of the no-vestibule condition.\\

The result presented in this appendix suggest that the vestibule improvement 
can be extrapolated to larger crowds. Nevertheless, we strongly advice 
exploring the scope and robustness of the vestibule in order to avoid unwanted 
consequences for the safety of evacuating pedestrians. \\

\section{\label{appendix_1}Flow vs. door size in the no-vestibule condition}

In Section \ref{results} we show the flow vs. $w$ relationships for the 1-door 
vestibule (Fig.~\ref{flow_vs_w_1door}) and the 2-doors vestibule 
(Fig.~\ref{flow_vs_w_2doors}). We noticed that the maximum flow value for the 
1-door vestibule is at $w = 6$, whereas the maximum flow value for 2-doors is 
at $w = 8$. In this appendix, we provide an explanation for this ``shifting'' 
behavior.\\ 

\begin{figure}[!htbp]
\centering
\subfloat[]{\includegraphics[width=0.49\columnwidth]
{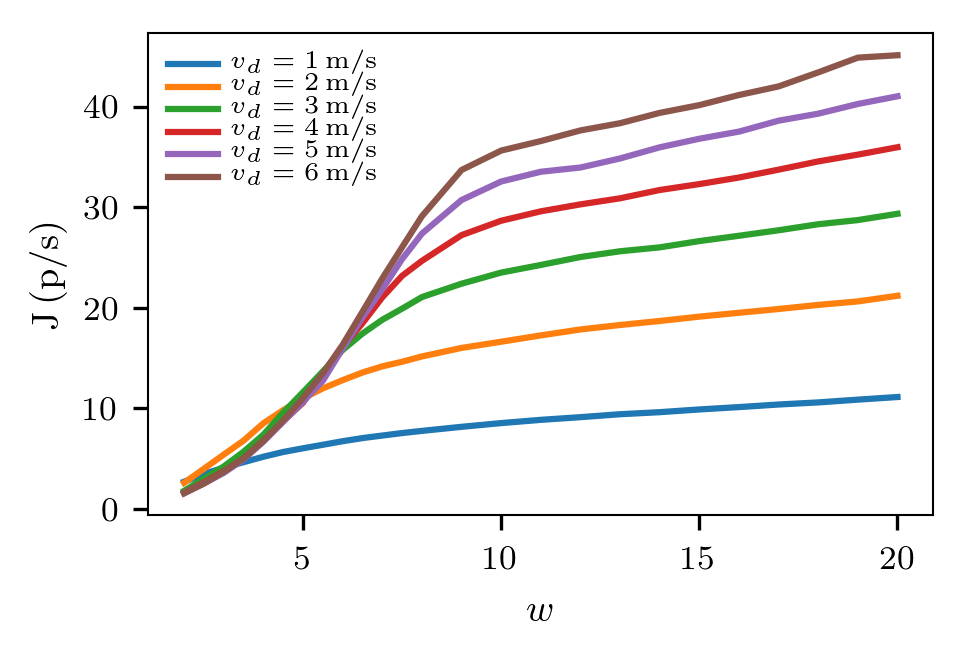}\label{flow_vs_w_all}}\ 
\subfloat[]{\includegraphics[width=0.49\columnwidth]
{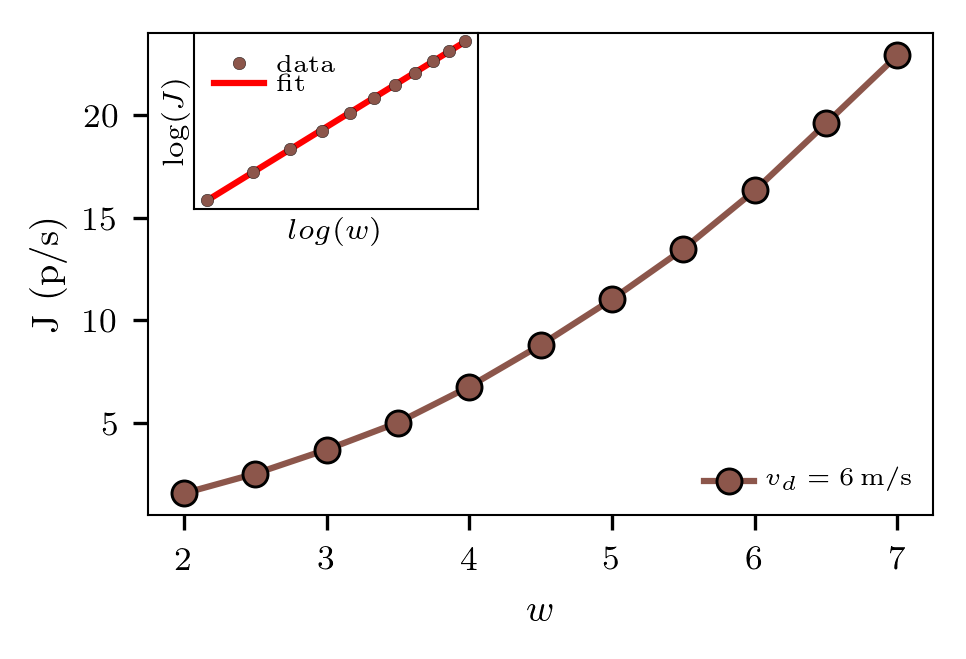}\label{flow_vs_w_vd6}}\\
\caption[width=0.47\columnwidth]{\textbf{(a)} Evacuation flow as a function of 
door width 
$w$ for the no-vestibule scenario. See the legend for the desired velocities 
explored. The initial crowd size is N=200. Data was averaged over 30 evacuation 
processes where the initial positions and velocities were random. The simulation 
finished when 180 pedestrians left the room. \textbf{(b)} Results corresponding 
to 
$v_d=6\,$m/s. The inset shows the linear regression applied to the log($J$) and 
log($w$). The regression is defined as $log(J) = m.log(w) + b$. The slope value 
obtained from the linear fit is $m=2.1$.}
\label{beverloo}
\end{figure}

First, we calculated the flow vs. $w$ in a regular bottleneck (no-vestibule 
scenario) to have a more fundamental understanding of the problem. 
Fig.~\ref{flow_vs_w_all} show the flow $J$ vs. the door width $w$ for 
different desired velocities. It is possible to observe 2 distinctive 
regimes.\\

The interval $w<7$ is characterized by the faster-is-slower. This regime is 
dominated by the blocking clusters and the friction force. On the other hand, 
the interval $w>7$ is characterized by the faster-is-faster phenomena (the 
higher $v_d$ the higher the evacuation flow). In this regime, the door is so 
large that the blocking clusters become unstable.\\

We focus on the faster-is-slower regime ($w<7$) and the desired velocities 
associated with high anxiety ($v_d \geq 3\,$m/s). Therefore, in 
Fig.~\ref{flow_vs_w_vd6} we ``zoom in'' and show the evacuation flow as a 
function of $w$ only for $v_d=6\,$m/s. It is worth mentioning that any $v_d\geq 
3 m/s$ exhibits a similar behavior. \\

At a first sight, it is possible to observe that the flow does not hold a linear 
relation in $w$. This result is in agreement with the laboratory and empirical 
results reported in Refs.~\cite{haghani2019push,gwynne2009questioning}. 
However, we acknowledge that there is an ongoing discussion on this 
topic since other experiments seem to yield a linear relation instead of a 
non-linear one~\cite{seyfried2009new,tian2012experimental}. \\

Nevertheless, the non-linear behavior exhibited in Fig.~\ref{flow_vs_w_vd6} is 
a sufficient justification for the ``shifting'' behavior in the flow vs. $w$ 
curves (Fig.~\ref{flow_vs_w_1door} and Fig.~\ref{flow_vs_w_2doors}). In other 
words, the non-linearity in $J(w)$ implies that a layout with only one exit 
of width $w$ yields more evacuation flow than a layout composed of 2 separated 
exits of $w/2$ each. If the $J(w)$ was linear, then we would expect no shifting 
behavior, since in this hypothetical case, the inflow to the vestibule 
provided by the 1-door scenario would be the same as the 2-doors scenario (at 
constant $w$). \\

In order to have a deeper understanding of the phenomena. We calculated a linear 
regression to the log data from Fig.~\ref{flow_vs_w_vd6} (see the inset plot). 
The slope obtained by this regression is $m=2.1$. This means that the flow 
vs. $w$ 
relation holds a quasi-quadratic behavior.\\ 

Although the analogy with granular media accelerated by gravity is not 
straightforward, we cannot fail to mention that there is an apparent 
similarity between the behavior described above and the Hagen-Beverloo's 
equation for the flow discharge of particles in a 
silo~\cite{beverloo1961flow}.\\

\section{\label{appendix_bc}Types of Blocking clusters}

In this appendix, we illustrate the blocking clusters that can be formed in the 
2-doors vestibule layout. Three types of blocking clusters (BC) can be 
distinguished. The ``exit door blocking cluster'', the ``inside vestibule 
blocking cluster'' (which is formed formed perpendicular to the vestibule 
length), and the ``vestibule door blocking cluster''. This latter type of BC is 
formed in the zone prior to the vestibule door.\\

\begin{figure}[!htbp]
\centering
{\includegraphics[width=0.7\columnwidth]
{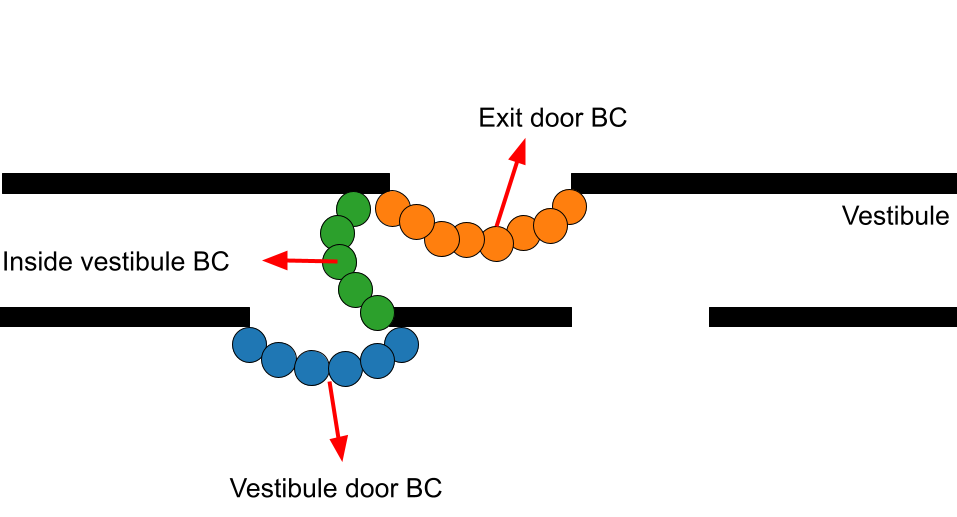}}\
\caption[width=0.49\columnwidth]{Picture of the blocking clusters in the 2-doors 
vestibule layout. Each blocking cluster type has a unique color (orange, green, 
and blue). The colors are in accordance with the curves from 
Fig.~\ref{pbc_2doors}. Those curves represent the blocking cluster probability 
as a function of the vestibule door width $w$.}
\label{bc_types}
\end{figure}

\section*{Acknowledgments}
This work wassupported by the Fondo para la investigaci\'on cient\'ifica  y 
tecnol\'ogica(FONCYT) grant Proyecto de investigaci\'on cient\'ifica y 
tecnol\'ogica Number PICT-2019-2019-01994. G.A.~Frank thanks Universidad 
Tecnológica Nacional (UTN) for partial support through Grant PID Number 
SIUTNBA0006595.  \\

\section*{References}

\bibliography{bibfile}

\end{document}